\documentclass[
  twocolumn,
  prb,
  amsmath,
  amssymb,
  superscriptaddress,
  floatfix
]{revtex4}

\usepackage{bm}
\usepackage{graphicx}
\usepackage{color}
\def \be {\begin{equation}}
\def \ee {\end{equation}}
\def \bea {\begin{align}}
\def \eea {\end{align}}

\def\bee{\begin{eqnarray}}
\def\eee{\end{eqnarray}}
\def \BC {\begin{cases}}
\def \EC {\end{cases}}

\begin{document}

\title{Quantum interference in HgTe structures}

\author{I.\ V.\ Gornyi}

\affiliation{
 Institut f\"ur Nanotechnologie, Karlsruhe Institute of Technology,
 76021 Karlsruhe, Germany
}
\affiliation{
 A.F.~Ioffe Physico-Technical Institute,
 194021 St.~Petersburg, Russia
}

\author{V. Yu. Kachorovskii}
\affiliation{
 A.F.~Ioffe Physico-Technical Institute,
 194021 St.~Petersburg, Russia
}

\author{A.\ D.\ Mirlin}

\affiliation{
 Institut f\"ur Nanotechnologie, Karlsruhe Institute of Technology,
 76021 Karlsruhe, Germany
}
\affiliation{
 Institut f\"ur Theorie der kondensierten Materie
 and
 DFG Center for Functional Nanostructures, Karlsruhe Institute of Technology,
 76128 Karlsruhe, Germany
}

\affiliation{
 Petersburg Nuclear Physics Institute,
 188300 St.~Petersburg, Russia.
}

\author{P.\ M.\ Ostrovsky}

\affiliation{Max-Planck-Institut f\"ur Festk\"orperforschung, Heisenbergstr. 1,
70569, Stuttgart, Germany}

\affiliation{
 L.~D.~Landau Institute for Theoretical Physics RAS,
 119334 Moscow, Russia
}

\begin{abstract}

We study quantum transport in HgTe/HgCdTe quantum
wells under the condition that the chemical potential is
located outside of the bandgap.
We first analyze symmetry properties of the effective Bernevig-Hughes-Zhang
Hamiltonian and the relevant symmetry-breaking perturbations.
Based on this analysis, we overview possible patterns of
symmetry breaking that govern the quantum interference (weak localization or
weak antilocalization) correction to the conductivity
in two dimensional HgTe/HgCdTe samples. Further, we perform a microscopic calculation of the quantum correction
beyond the diffusion approximation. Finally, the interference
correction and the low-field magnetoresistance in a quasi-one-dimensional
geometry are analyzed.

\end{abstract}

\maketitle

\section{Introduction}
\label{s1}

Two-dimensional (2D) and three-dimensional (3D) materials and structures with
strong spin-orbit interaction in the absence of magnetic field (i.e. with
preserved time-reversal invariance) may exhibit a topological insulator (TI)
phase \cite{Hasan10RMP,Qi11,kane,BernevigHughesZhang,Koenig07,Fu07,hasan}.
In the 2D case, the TI behavior was experimentally discovered by the W{\"u}rzburg group \cite{Koenig07} in
HgTe/HgCdTe quantum wells (QWs) with band gap inversion due to strong spin-orbit interaction.
The band inversion results in emergence of helical modes at the edge of the sample.
These modes are topologically protected as long as the time-reversal symmetry is preserved.

Application of a bias voltage leads to the appearance of a
quantized transverse spin current, which is the essence of the quantum spin-Hall
effect (QSHE). An interplay between the charge and spin degrees of freedom
characteristic to QSHE is promising for the spintronic applications.
The existence of delocalized mode at the edge of an inverted 2D HgTe/HgCdTe
QW was demonstrated in Refs. \onlinecite{Koenig07,Roth09}.
These experiments have shown that HgTe/HgCdTe structures realize
a novel remarkable class of materials---$Z_2$ topological insulators---and thus opened
a new exciting research direction. Another realization of a 2D $Z_2$ TI based on
InAs/GaSb structures proposed in Ref.~\onlinecite{Liu08} was experimentally
discovered in Ref. \onlinecite{Du}.

When the chemical potential is shifted by applying a gate voltage away from the band gap,
a HgTe/HgCdTe QW realizes a 2D metallic state which can be termed a 2D spin-orbit
metal. The interference corrections to the conductivity and the low-field magnetoconductivity
of such a system reflect the Dirac-fermion character of carriers
\cite{ostrovsky10,Tkachov11,OGM12,Richter12}, similarly to
interference phenomena in graphene\cite{McCann,Nestoklon,Ostrovsky06,aleiner-efetov}
and in surface layers of 3D TI \cite{ostrovsky10,Glazman,Koenig13}.
Recently, the magnetoresistivity of HgTe/HgCdTe structures was experimentally
studied away from the insulating regime in Refs. \onlinecite{kvon,minkov,bruene},
both for inverted and normal band ordering.

In this article we present a systematic theory of
the interference-induced quantum corrections to the conductivity
of HgTe-based structures in the metallic regime.
We investigate the quantum interference in the whole spectrum,
from the range of almost linear dispersion to the vicinity of the band bottom
and address the crossover between the regimes.
We begin by analyzing
in Sec.~\ref{s2} symmetry properties of the underlying Dirac-type Hamiltonian
and physically important symmetry-breaking mechanisms.
In Sec.~\ref{s3} we overview a general symmetry-based approach~\cite{OGM12} to the problem
and employ it to evaluate the conductivity corrections
within the diffusion approximation.
Section~\ref{s4} complements the symmetry-based
analysis by microscopic calculations. Specifically, we calculate
the interference correction beyond the diffusion approximation,
by using the kinetic equation for Cooperon modes which includes
all the ballistic effects.
A quasi one-dimensional geometry is analyzed in Sec.~\ref{s5}.
Section \ref{s6} summarizes our results and
discusses a connection to experimental works.

\section{Symmetry analysis of the low-energy Hamiltonian}
\label{s2}

The low-energy Hamiltonian for a symmetric HgTe/HgCdTe structure was introduced in
Ref.\ \onlinecite{BernevigHughesZhang} in the framework of the
$\mathbf{k}\cdot\mathbf{p}$ method. The Bernevig-Hughes-Zhang (BHZ)
Hamiltonian possesses a $4 \times 4$ matrix
structure in the Kramers-partner space
and E1 -- H1 space of electron- and hole-type levels~\cite{Qi11,Liu08,Rothe10},
\begin{gather}
 H_\text{BHZ}
  = \begin{pmatrix}
      h(\mathbf{k}) & 0 \\
      0 & h^*(-\mathbf{k})
    \end{pmatrix}, \\
 h(\mathbf{k})
  = \begin{pmatrix}
      \epsilon(\mathbf{k}) + m(\mathbf{k}) & A (k_x + i k_y) \\
      A (k_x - i k_y) & \epsilon(\mathbf{k}) - m(\mathbf{k})
    \end{pmatrix}.
    \label{1}
\end{gather}
Here the components of spinors are ordered as $E1+,H1+,E1-,H1-$.
It is convenient to introduce Pauli matrices $\sigma_{0,x,y,z}$ for the E1---H1 space
and $s_{0,x,y,z}$ for the Kramers-partner space (here $\sigma_0$ and $s_0$ are unity matrices),
yielding
\begin{equation}
H_\text{BHZ}
  = \epsilon(\mathbf{k})\sigma_0 s_0 + m(\mathbf{k}) \sigma_z s_0 + A
 k_x \sigma_x s_z - A k_y \sigma_y s_0.
\label{2}
\end{equation}

The effective  mass $m(\mathbf{k})$  and energy $\epsilon(\mathbf{k})$ are given by
\begin{equation}
 m(\mathbf{k})
  = M + B \mathbf{k}^2,
  \qquad
   \epsilon(\mathbf{k})
  = C + D \mathbf{k}^2,
  \label{3}
\end{equation}
The normal insulator phase corresponds to $M>0$ (which is realized in thin QWs, $d<6.2$ nm),
whereas the TI phase is characterized by $M<0$ (realized in thick QW).~\cite{Koenig07}

The Hamiltonian $H_\text{BHZ}$ breaks up into two blocks
that have the same spectrum
\begin{equation}
 E^\pm_\mathbf{k}
  = \epsilon(\mathbf{k}) \pm \sqrt{A^2 k^2 + m^2(\mathbf{k})}.
  \label{4}
\end{equation}
The eigenfunctions for each block are two-component spinors in E1-H1 space:
\be
\psi_\mathbf{k}^{(\pm)}(\mathbf{r})=\chi_\mathbf{k}^{(\pm)}\, e^{i\mathbf{k r}}
\ee
where spinors $\chi_\mathbf{k}^{(\pm)}$ are different in different blocks
\begin{align}
\chi_\mathbf{k}^{(\text{I},\pm)}&=
(1+\mu^2_\pm)^{-1/2}\
\begin{pmatrix}
1\\
\mu_{\pm} e^{-i \phi_\mathbf{k}}
\end{pmatrix},
\\
\chi_\mathbf{k}^{(\text{II},\pm)}&= (1+\mu^2_\pm)^{-1/2}\
\begin{pmatrix}
1\\
-\mu_{\pm} e^{i \phi_\mathbf{k}}.
\end{pmatrix}
\end{align}
Here $\phi_\mathbf{\bf k}$ is the polar angle of the momentum $\mathbf{k}$
and
\begin{equation}
\mu_{\pm}=\frac{\pm \left[A^2k^2+m^2(\mathbf{k})\right]^{1/2}-m(\mathbf{k})}{A k}
\end{equation}
corresponds to the upper and lower branches $E^\pm_\mathbf{k}$ of the spectrum.

Disorder potential $V(\mathbf{r})$ is conventionally introduced in the BHZ model by adding the
scalar term~\cite{Tkachov11}
\begin{equation}
 H_\text{dis}
  = V(\mathbf{r}) \sigma_0 s_0
  \label{5}
\end{equation}
to the $4\times 4$ Hamiltonian $H_\text{BHZ}$.
This model describes smooth disorder that does not break the spatial reflection symmetry of the structure and thus
does not mix the two Kramers blocks of the BHZ Hamiltonian.

We are now going to discuss symmetry properties of the Hamiltonian
of a 2D HgTe/CdTe QW and symmetry-breaking mechanisms \cite{Rothe10,OGM12}.
The Hamiltonian $H_\text{BHZ} + H_\text{dis}$ is characterized by
the exact global time-reversal (TR) symmetry $H(\mathbf{k}) = s_y H^*(-\mathbf{k}) s_y$.
Further, this Hamiltonian commutes with $s_z$, which we term the ``spin symmetry''.
Finally, an additional approximate symmetry operative within each Kramers block
emerges for some regions of energy. Specifically,
$h(\mathbf{k})$ in the inverted regime acquires the exact
symplectic ``block-wise'' TR symmetry when $m(k_F)=M+Bk_F^2=0$. Around this point, the symmetry is
approximate.  An approximate orthogonal block-wise TR symmetry emerges
near the band bottom for $|M| \gg \{A k_F, B k_F^2\}$ and for high energies
$B k_F^2 \gg \{|M|, A k_F\}$.

When employing the symmetry analysis to a realistic system, the symmetries of Hamiltonian (\ref{2})
should be regarded as approximate.
The term ``approximate symmetry'' here means that the corresponding
symmetry breaking perturbations in the Hamiltonian are weak, such that they violate
this ``approximate symmetry'' on scales that are much larger than the mean free path.
On the technical level, the gaps of the corresponding soft modes (Cooperons) are small
in this case. This, in turn, implies that there exists an intermediate regime, when the dephasing
length (or the system size) is shorter than the corresponding symmetry-breaking length.
In this regime, the diffusive logarithmic correction
to the conductivity is insensitive to this symmetry-breaking mechanism and the system behaves
as if this symmetry is exact. However, when the dephasing length becomes longer than
the symmetry-breaking length, the relevant singular corrections are no longer determined
by the dephasing but are cut off by the symmetry-breaking scale.
This signifies a crossover to a different (approximate) symmetry class.
Below we analyze the relevant symmetry-breaking perturbations in HgTe structures.

The $s_z$ \textit{spin symmetry} is violated by perturbations that do not preserve
the reflection ($z\to -z$) symmetry of the QW.
Such perturbations yield nonzero block-off-diagonal elements in the full $4\times 4$ low-energy
Hamiltonian. One of possible sources for the block mixing is
the bulk inversion asymmetry (BIA) of the
HgTe lattice. The corresponding term in the effective Hamiltonian reads \cite{Liu08}
\begin{equation}
 H_\text{BIA}
  = \begin{pmatrix}
      0 & 0 & \delta_e k_+ & -\Delta_0 \\
      0 & 0 & \Delta_0 & \delta_h k_- \\
      \delta_e k_- & \Delta_0 & 0 & 0 \\
      -\Delta_0 & \delta_h k_+ & 0 & 0
    \end{pmatrix}.
  \label{6}
\end{equation}
The BIA perturbation (\ref{6}) contains the momentum-independent term
with $\Delta_0$ that connects the electronic and heavy-hole bands~\cite{Winkler}
with opposite spin projections. The terms with $\delta_e$ and $\delta_h$ stem
from the cubic Dresselhaus spin-orbit interaction within $\Gamma_6$ and $\Gamma_8$, respectively.
Further, the $s_z$ symmetry is broken by the Rashba spin-orbit interaction due to the structural inversion
asymmetry (SIA):\cite{Liu08,Rothe10}
\begin{equation}
H_R=\begin{pmatrix}
      0 & 0 &  i r_e k_- & 0 \\
      0 & 0 & 0 & 0 \\
      -  i r_e k_+ & 0 & 0 & 0 \\
      0 & 0 & 0 & 0
    \end{pmatrix},
\label{8}
\end{equation}
Here only the linear-in-momentum E1 SIA term is retained, as
the SIA terms for heavy holes contain higher powers of $k$.
Finally, short-range impurities and defects, as well as HgTe/HgCdTe interface roughness
may also violate the $z\to -z$ symmetry of the QW, giving rise to
a random local block-off-diagonal perturbations.

\section{Symmetry analysis of quantum conductivity corrections}
\label{s3}

Here we overview the approach developed in Ref. \onlinecite{OGM12} for the analysis
of quantum-interference corrections to the conductivity of an infinite 2D HgTe QW.
Within the diffusion approximation, conductivity
corrections that are logarithmic in temperature $T$
are associated with certain TR symmetries.
The TR symmetry transformations can be represented as anti-unitary operators
that act on a given operator $\mathcal{O}$ according to
\begin{equation*}
T: \quad \mathcal{O}
  \mapsto U^{-1} \mathcal{O}^T U.
\end{equation*}
Here $U$ is some unitary operator (note that
the momentum operator changes sign under transposition).

When the Hamiltonian of the system is given by a $4\times 4$ matrix,
possible TR symmetry transformations can be cast in the form involving the tensor products
of Pauli matrices:
\begin{equation}
T_{ij}: \quad \mathcal{O}
  \mapsto \sigma_i s_j \mathcal{O}^T s_j \sigma_i, \quad i,j=0,x,y,z.
  \label{12}
\end{equation}
Each of these TR symmetries corresponds to a Cooperon mode contributing to the
singular  one-loop conductivity correction:
\begin{equation}
\delta \sigma_{ij} = - c_{ij} \frac{e^2}{2 \pi h}
\ln\left(\frac{\tau_\phi}{\tau}\right), \qquad c_{ij}=-1,\ 0,\ 1.
\label{deltasigmasym}
\end{equation}
Here $\tau_\phi$ is the phase-breaking time due to inelastic scattering and
$\tau$ is the transport time.
The factors $c_{ij}$ in Eq.~(\ref{deltasigmasym}) are zero when the TR symmetry $T_{ij}$ is broken by the Hamiltonian;
otherwise, $c_{ij}=T_{ij}^2=\pm 1$  for the orthogonal and symplectic type of
the TR symmetry, respectively.
The above perturbative loop expansion is justified by the large parameter $E_F \tau\gg 1$,
where $E_F$ is the Fermi energy counted from the bottom of the band.

An analogous symmetry analysis of the interference effects was performed for a related problem of
massless Dirac fermions in graphene in Ref. \onlinecite{Ostrovsky06}.
In Ref. \onlinecite{OGM12} this approach was generalized to the case of massive Dirac fermions
in a HgTe QW. By choosing the basis H1+,E1+,E1-,H1-, the linear-in-$k$
term in the BHZ Hamiltonian acquires the same structure as in Ref. \onlinecite{Ostrovsky06}:
\begin{equation}
H_{BHZ}=\epsilon(\mathbf{k})\sigma_0 s_0
+ [-m(\mathbf{k}) \sigma_z + A \mathbf{k} {\boldsymbol\sigma}] s_z.
\label{9}
\end{equation}

When the chemical potential is located in the range of approximately linear spectrum,
$E_\pm \simeq \pm A |k|$, the Dirac mass
$H_M=-m(\mathbf{k})\sigma_z s_z$
and the $s_z$-symmetry breaking terms
\begin{align}
H_\text{BIA}
  &= \Delta_0 \sigma_z s_x + \delta_+ (k_x\sigma_x+k_y\sigma_y) s_x
  \nonumber \\
  &\qquad
  +  \delta_-(k_x\sigma_y-k_y\sigma_x) s_y,
  \label{10}
  \\
H_R &= (r_e/2)[(k_x\sigma_y+k_y\sigma_x)s_x-(k_x\sigma_x-k_y\sigma_y)s_y],
\label{11}
\end{align}
[where $\delta_\pm=(\delta_h\pm\delta_e)/2$]
can be treated as weak perturbations to the massless (graphene-like) Dirac Hamiltonian:
\begin{equation}
H_A=A(k_x\sigma_x+k_y\sigma_y)s_z.
\label{15}
\end{equation}
The latter possesses four TR symmetries:
\begin{align}
 T_{xx}:&\quad \mathcal{O}
  \mapsto \sigma_x s_x \mathcal{O}^T \sigma_x s_x,\qquad T_{xx}^2=1,
  \label{16a} \\
 T_{y0}:&\quad \mathcal{O}
  \mapsto \sigma_y s_0 \mathcal{O}^T \sigma_y s_0,\qquad T_{y0}^2=-1,
  \label{16b} \\
 T_{yz}:&\quad \mathcal{O}
  \mapsto \sigma_y s_z \mathcal{O}^T \sigma_y s_z,\qquad T_{yz}^2=-1,
  \label{16c} \\
 T_{xy}:&\quad \mathcal{O}
  \mapsto \sigma_x s_y \mathcal{O}^T \sigma_x s_y,\qquad T_{xy}^2=-1.
  \label{16d}
\end{align}
These symmetries give rise to a positive weak antilocalization (WAL) conductivity correction
\begin{equation}
\delta\sigma=-(1-3)\frac{e^2}{2 \pi h} \ln\left(\frac{\tau_\phi}{\tau}\right)
=2\times \frac{e^2}{2 \pi h} \ln\left(\frac{\tau_\phi}{\tau}\right),
\label{2WAL}
\end{equation}
corresponding to two independent copies of a symplectic-class system (2Sp).

The mass term $H_M=-m(\mathbf{k})\sigma_zs_z$ violates $T_{y0}$ and $T_{yz}$ symmetries~\cite{footnote_mass}
on the scale determined by the symmetry breaking rate~\cite{Tkachov11,OGM12}
$1/\tau_m\sim [m(k_F)/E_F]^2/\tau$ (see Sec. \ref{s4} below for the microscopic derivation).
The two out of four soft modes acquire the gap $\tau/\tau_m$, yielding
\begin{equation}
 \delta\sigma
  =
-2\times \frac{e^2}{2 \pi h}\ln\left(
      \frac{\tau}{\tau_\phi}+\frac{\tau}{\tau_m}  \right).
\end{equation}
At lowest temperatures, when $\tau_\phi\gg\tau_m$, we find a nonsingular-in-$T$ result:
\begin{equation}
 \delta\sigma
  \simeq 2\times \frac{e^2}{2 \pi h}\ln\left(\frac{\tau_m}{\tau}  \right).
  \label{2U}
\end{equation}
For higher temperatures, when $\tau_\phi\ll \tau_m$, these two copies of
a unitary-class system (2U) become
two copies of the (approximately) symplectic class, with the correction given by
Eq. (\ref{2WAL}).

In the presence of inversion-asymmetry terms $H_\text{BIA}$ and $H_R$, the only remaining TR symmetry is
$T_{xy}$. The symmetry analysis yields the following
expression for the conductivity correction in this (generic) case \cite{OGM12}:
\begin{multline}
 \delta\sigma
  = \frac{e^2}{2 \pi h} \Bigg[
  \ln\left( \frac{\tau}{\tau_\phi} + \frac{\tau}{\tau_\Delta}  +
\frac{\tau}{\tau_\text{SO}} \right) \\
  -\ln\left(
        \frac{\tau}{\tau_\phi} + \frac{\tau}{\tau_m} +
        \frac{\tau}{\tau_\Delta} + \frac{\tau}{\tau_\text{SO}}
      \right) \\
            -\ln\left( \frac{\tau}{\tau_\phi} + \frac{\tau}{\tau_m} +
              \frac{\tau}{\tau_\text{SO}} \right)
            -\ln\frac{\tau}{\tau_\phi}
    \Bigg].
    \label{SpFull}
\end{multline}
Here $1/\tau_\Delta$ is the symmetry-breaking rate due to the
$k$-independent term $\Delta_0\sigma_zs_x$ in $H_\text{BIA}$ while
$1/\tau_{SO}$ describes the $s_z$-symmetry breaking governed by
linear-in-$k$ terms in $H_\text{BIA}$ and $H_R$.

Thus the behavior of the conductivity at the lowest $T$ is governed
by the single soft mode which reflects the physical symplectic TR symmetry
$T_{xy}$.
This mode yields a WAL correction characteristic for a single copy of
the symplectic class system (1Sp).
At higher temperatures, depending on the hierarchy of symmetry-breaking rates,
the folllowing patterns of symmetry breaking can be realized:~\cite{OGM12}
\begin{itemize}
\item \quad $\tau \ll\tau_m\ll\text{min}[\tau_\Delta,\tau_{SO}]$:  2Sp
$\to$ 2U $\to$ 1Sp;
\item \quad $\tau \ll\text{min}[\tau_\Delta,\tau_{SO}]\ll\tau_m$: 2Sp $\to$ 1Sp.
\item \quad $\tau_\text{SO}\alt \tau$ or $\text{max}[\tau_m,\tau_\Delta]\alt\tau$: 1Sp.
\end{itemize}

We now turn to the case when the chemical potential is located in the bottom
of the spectrum, $E_F\ll |m|$. In this limit, the spectrum is approximately parabolic:
\begin{eqnarray}
 E_+(\mathbf{k})
  &\simeq& |M| + B k^2 +
    \epsilon(\mathbf{k})+A^2 k^2/|M|.
  \label{spectrum}
\end{eqnarray}
The direction of the pseudospin within each block is almost frozen by the
effective ``Zeeman term'' $M$. The linear-in-$k$ terms of the BHZ Hamiltonian
can then be treated as a weak spin-orbit-like perturbation to the massive diagonal Hamiltonian
\begin{equation}
H_M= - m(\mathbf{k}) \sigma_z  s_z.
\label{HamMass}
\end{equation}
Neglecting the block mixing, the conductivity is given by a sum
of two weak localization (WL) corrections characteristic for an orthogonal symmetry class:
\begin{equation}
\delta \sigma=2\times \frac{e^2}{2\pi h}
\ln\left(\frac{\tau}{\tau_A}+\frac{\tau}{\tau_\phi}\right).
\label{2O2U}
\end{equation}
Here $1/\tau_A\sim [A k_F/m(k_F)]^4/\tau$ is the symmetry-breaking rate due to ``relativistic'' correction
$H_A$.
The microscopic derivation of $1/\tau_A$ is performed in Sec. \ref{s4} below.

The TR symmetries of the Hamiltonian $H_M$ can be combined into four pairs:
\begin{equation*}
\begin{aligned}
 T_{00} \sim T_{zz}, \quad T_{0z} \sim T_{z0},
 \quad
 T_{xx} \sim T_{yy}, \quad T_{xy} \sim T_{yx}.
 \end{aligned}
\end{equation*}
Symmetry breaking perturbations can affect the symmetries from each
pair in different ways. When both TR symmetries from the pair are respected by
the
perturbation, the full Hamiltonian decouples into two blocks corresponding to
the eigenvalues of
$\sigma_z s_z$. Such a pair contributes to the conductivity as if there is a
single TR symmetry. If only one of the two TR symmetries is broken within the
pair, the remaining symmetry yields a conventional singular contribution.
Finally, when both symmetries within the pair are broken, such pair does not
contribute.

Thus, when both symmetries are not simultaneously violated, each pair
contributes as a single soft mode. Note that in this case the corresponding Cooperon
mass is determined by the sum of symmetry-breaking times rather than by the
sum of rates.
Following this rule, the inclusion of $H_A$, $H_\text{BIA}$, and $H_R$
gives rise to the following interference correction:~\cite{OGM12}
\begin{multline}
 \delta\sigma
   = \frac{e^2}{2 \pi h}\Bigg[2 \ln\left(
       \frac{\tau}{\tau_\phi}+\frac{\tau}{\tau_A} +\frac{\tau}{\tau_{SO}}
     \right) \\
     +\ln\left( \frac{\tau}{\tau_\phi}+\frac{\tau}{\tau_{SO}}
     +\frac{\tau}{\tau_A +\tau_\Delta}\right)
     -\ln \frac{\tau}{\tau_\phi} \Bigg].
     \label{OFull}
\end{multline}
The only true massless mode in Eq.~(\ref{OFull}) stems again from the physical
symplectic TR symmetry $T_{xy}$.
This means that the  generic block-mixing terms drive the two copies of
the (approximately) orthogonal class
to a single copy of a symplectic-class system. The hierarchy of the
symmetry-breaking rates $\tau_A^{-1}$,
$\tau_\Delta^{-1}$, and $\tau_{SO}^{-1}$, generates the following three patterns of crossovers:\cite{OGM12}
\begin{itemize}
\item\quad $\tau\ll\tau_A\ll\text{min}[\tau_\Delta, \tau_{SO}]$: 2O $\to$ 2U $\to$ 1Sp.
\item\quad $\text{min}[\tau_\Delta, \tau_{SO}]\ll \tau_A $ and $\tau\ll \text{min}[\tau_A, \tau_{SO}]$:\\ \phantom{aa} 2O $\to$ 1Sp.
\item\quad $\tau_{SO} \alt \tau$: 1Sp.
\end{itemize}

%%%%%%%%%%%%%%%%%%%%%%%%%%%%%%%%%%%%%%%%%%%%%%%%%%%%%%%%
\begin{figure*}
 \includegraphics[width=0.46\textwidth]{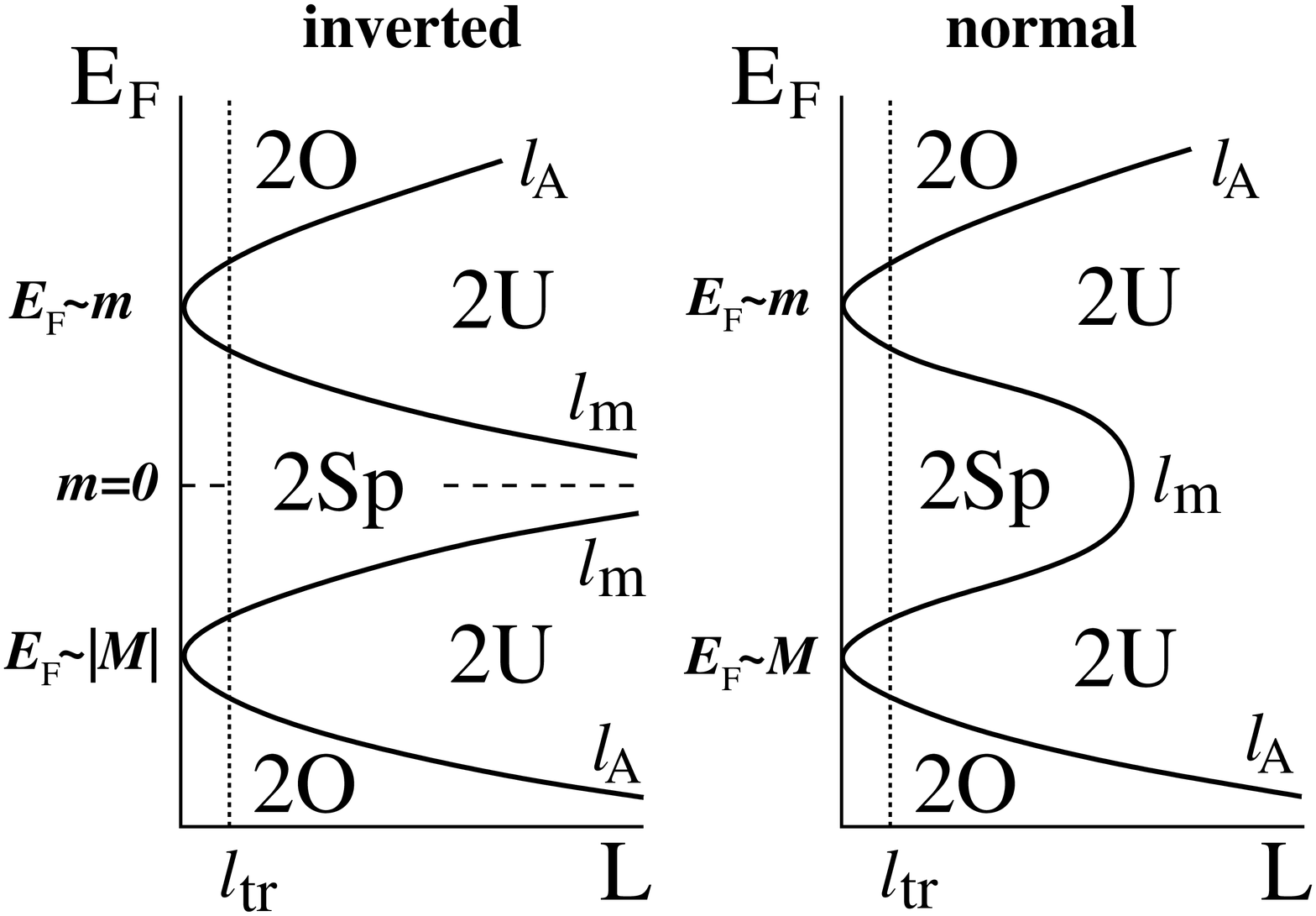}\hspace*{0.2cm}
\includegraphics[width=0.39\textwidth]{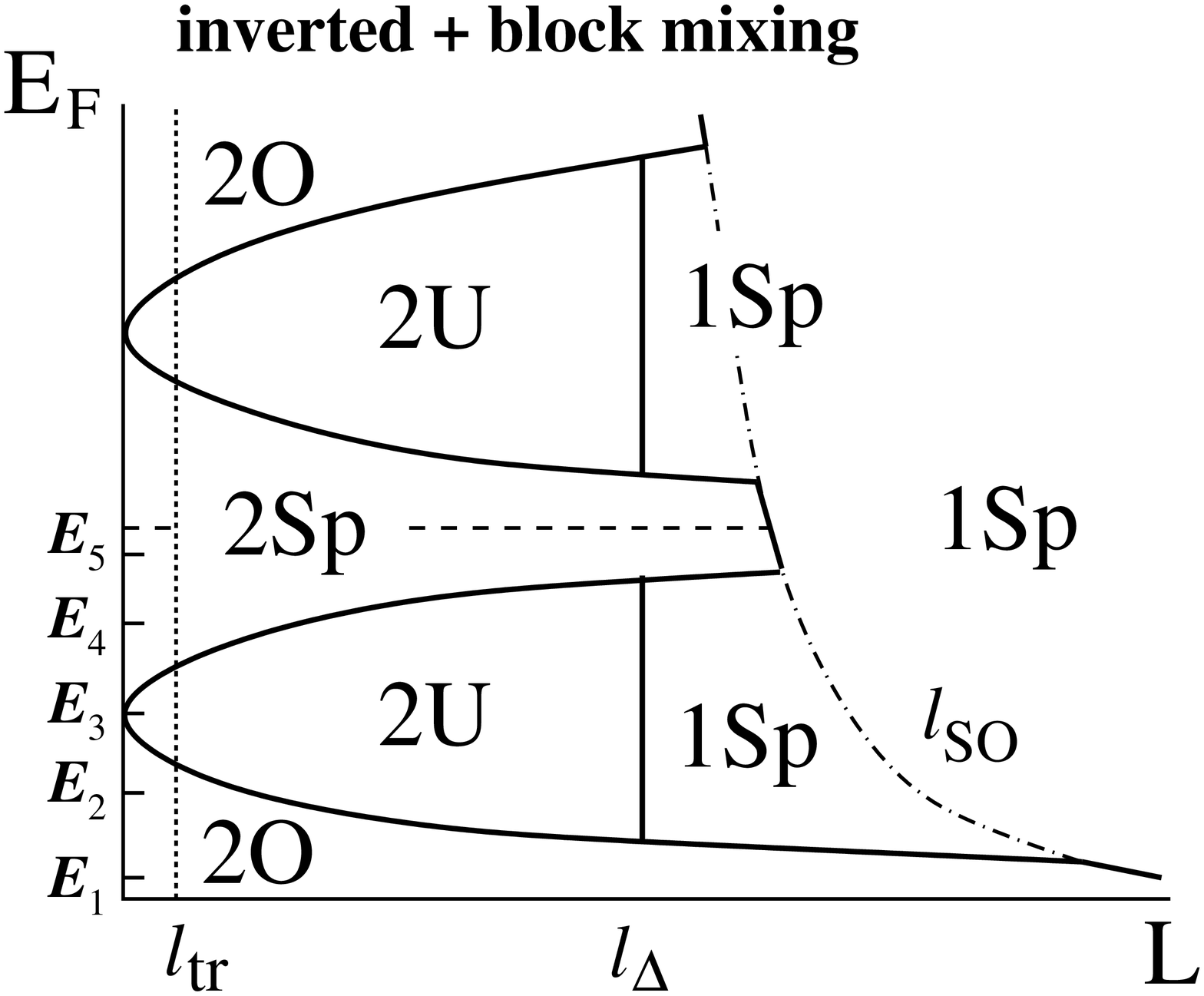}
 \caption{``Phase diagrams'' showing the symmetry patterns for the quantum correction
 to the conductivity in a 2D HgTe QW when the chemical potential is located away from the bandgap, $E_F>0$.
The length $L$ is the smallest of the system size or the dephasing length $L_\phi$.
The transport scattering length $l_\text{tr}$ (shown by a vertical dotted line)
and the BIA-splitting length $l_\Delta=(D\tau_\Delta)^{1/2}$ ($D$ is the diffusion constant)
are assumed to be independent of energy.
The ``phase boundaries'' (solid lines) of the 2U-regions are defined by
$l_m=(D\tau_{m})^{1/2}$  for 2Sp $\to$ 2U crossover, and by $l_A=(D\tau_{A})^{1/2}$
for 2O $\to$ 2U crossover.
 \textit{Left panel}: Inverted band structure (thick quantum well) with no
block mixing. Dashed line shows energy for which $m(k_F)=-|M|+Bk_F^2=0.$
 \textit{Middle panel}: Normal band structure (thin quantum well) with no block
mixing.
 \textit{Right panel}: Inverted band structure with block mixing
 characterized by $l_\Delta=(D\tau_\Delta)^{1/2}$ and
$l_\text{SO}=(D\tau_{SO})^{1/2}$ (dash-dotted),
with $\tau<\tau_\Delta<\tau_{SO}$.
The energies $E_F=E_1\ldots E_5$ (from bottom to top) mark different horizontal
cross-sections
of the ``phase-diagram'' corresponding to the patterns 2O $\to$ 1Sp, 2O $\to$
2U $\to$ 1Sp,
2U $\to$ 1Sp, 2Sp $\to$ 2U $\to$ 1Sp, and 2Sp $\to$ 1Sp, respectively,
that appear with increasing $L$.
The perturbative one-loop results discussed in Sec.~\ref{s3} require the condition $E_F>1/\tau$, which
introduces an additional horizontal line near the band bottom in all the panels. Further, we assume that the localization
length in the 2U and 2O regions is smaller than the scale $L$.}
 \label{Fig1}
\end{figure*}
%%%%%%%%%%%%%%%%%%%%%%%%%%%%%%%%%%%%%%%%%%%%%%%%%%%%%%%%%%%%%%%%%%%%%%%%%%%%%%

To summarize this section, we have analyzed the quantum conductivity
correction in the diffusion approximation using the symmetry-based approach.
We have identified various possible types of behavior that
include 2O, 2U, 2Sp, and 1Sp regimes. The $T$-dependence of the conductivity
correction
is given by $\delta\sigma=\alpha (e^2/\pi h) \ln \tau_\phi(T)$, where $\alpha = -1, 0, 1$, and $1/2$, respectively.
The ``phase diagram'' describing these regimes is shown in Fig. \ref{Fig1}.

In general, crossovers between the regimes are governed by four symmetry
breaking rates:
$1/\tau_{SO},$ $1/\tau_\Delta,$ $1/\tau_m,$ and $1/\tau_A$.
The first two describe a weak block mixing in the BHZ Hamiltonian.
They are present for arbitrary position of the Fermi energy
and are assumed to be smaller than $1/\tau$.
Near the band bottom (and for very high energies, where the spectrum is no longer linear)
the ``intra-block'' rates satisfy: $\tau/\tau_m\sim 1$, while $\tau/\tau_A\ll 1$.
In the region of linear spectrum the relations are opposite:
 $\tau/\tau_m\ll 1$, while $\tau/\tau_A\sim 1$.

Assuming for simplicity the absence of the BIA splitting of the spectrum, $1/\tau_\Delta=0,$
the general expression for the conductivity correction can be written as:
\begin{eqnarray}
\delta\sigma &=&
\frac{e^2}{4\pi^2 \hbar}
\times \left[2\ln\left(\frac{\tau}{\tau_\phi}+\frac{\tau}{\tau_A}+\frac{\tau}{\tau_{SO}}\right)
\right.\nonumber \\
&-&
\left.
2\ln\left(\frac{\tau}{\tau_\phi}+\frac{\tau}{\tau_m}+\frac{\tau}{\tau_{SO}}\right)
\right.\nonumber \\
&+& \left. \ln\left(\frac{\tau}{\tau_\phi}+\frac{\tau}{\tau_{SO}}\right)
-\ln\left(\frac{\tau}{\tau_\phi}\right)
\right].
\label{interpol}
\end{eqnarray}
The first term here describes two copies (decoupled blocks) of WL near the
band bottom,
the second term describes two copies (decoupled blocks) of WAL in the range
of linear dispersion,
and the last two terms reflect a block mixing due to the spin-orbit
interaction/scattering
(they are present at any energy).

\section{Microscopic calculation of the interference correction}
\label{s4}

In this section, we present a microscopic calculation of the interference correction
to the conductivity for white-noise disorder beyond the diffusive approximation.
We first consider the model with decoupled blocks and later analyze the effect of block mixing.

The WAL correction for decoupled blocks was studied in Ref. \onlinecite{Tkachov11}
within the diffusive approximation for the case
when the chemical potential is located in the almost linear range of the spectrum.
It was shown there that the finite bandgap (leading to a weak nonlinearity of dispersion)
suppresses the quantum interference on large scales. Here we calculate the interference-induced
conductivity correction in the whole range of concentrations and without relying on the diffusion
approximation. This allows us to describe analytically the crossover from the WL behavior near the
band bottom to the WAL in the range of almost linear spectrum.
We compare our results to those of Ref. \onlinecite{Tkachov11} in the end of Sec.~\ref{Corr}.

For simplicity, we will consider the case $B=C=D=0$.
Then the two blocks of the BHZ Hamiltonian read:
\begin{eqnarray}
H_\text{I}&=&\left[
                    \begin{array}{cc}
                     M & A (k_x + i k_y)  \\
      A (k_x - i k_y) & -M \\
                    \end{array}
                  \right],
                  \\
                   H_\text{II}&=&\left[
                    \begin{array}{cc}
                      M & -A (k_x - i k_y)  \\
      -A (k_x + i k_y) & -M \\
                    \end{array}
                  \right].
\label{I,II}
\end{eqnarray}
A generalization onto the case of $k$-dependent mass $m(\textbf{k})$
is straightforward. For definiteness, we will consider the block $H_\text{II}$.

The bare Green's function of the system is a $2\times2$ matrix in E1-H1 space
which can be represented as a sum of the contributions of upper and lower branches:
\begin{equation}
\hat{G}(E,\mathbf{p})=
\frac{\hat{P}_+(\mathbf{k})}{E-E^+_\mathbf{k}}+\frac{\hat{P}_-(\mathbf{k})}{E-E^-_\mathbf{k}},
\end{equation}
where the projectors $\hat{P}_{\pm}$ are given by
\begin{equation}
\hat{P}_{\pm}(\mathbf{k})=|\chi^{(\pm)}_\mathbf{k}\rangle\,\langle\chi^{(\pm)}_\mathbf{k}|.
\end{equation}

Making use of the condition $k_F l\gg 1$, we can neglect the contribution of the lower branch when considering
the interference corrections for $E_F$ residing in the upper band,
\be
E_\mathbf k^{+} =\sqrt{M^2+A^2k^2}.
\ee
This allows us to retain in the matrix Green's function only the contribution of the upper band:
\begin{eqnarray}
\hat{G}(E,\mathbf{k})&\simeq&
\frac{\hat{P}_+(\mathbf{k})}{E-E^+_\mathbf{k}-\Sigma_+}
=\hat{P}_+(\mathbf{k})G_+(E,\mathbf{k}),
\end{eqnarray}
where $\Sigma_+$ is the disorder-induced self-energy.
From now on we will omit the branch index ``+''.
The spinors in the upper band of block II read:
\begin{align}
\chi_\mathbf{k} &=
\frac{1}{\sqrt{1+\mu^2}}\
\begin{pmatrix}
1\\
-\mu e^{i \phi_\mathbf{k}}
\end{pmatrix},\label{chi1}
\\
\mu & =\frac{ A k}{ M + \sqrt{M^2+ A^2 k^2}}.
\label{mu-def}
\end{align}

While the diffusive behavior of the quantum interference
correction is universal, the precise from of the correction in the ballistic regime depends
on the particular form of the disorder correlation function.
In what follows, we will assume a white-noise correlated disorder
with
\begin{equation}
\langle V(\mathbf{r}) V(\mathbf{r}') \rangle = W_0 \delta(\mathbf{r}-\mathbf{r}').
\label{corrW}
\end{equation}
Within this model the crossover between the diffusive and ballistic regimes can be
described analytically.

Next, we notice that in the  standard diagrammatic technique, each impurity vertex
$V(\mathbf{k}-\mathbf{k}')$ is sandwiched between two ``projected''
Green's functions. Therefore, we can dress the impurity vertices by adjacent
parts of the projectors, thus replacing in all diagrams
$$\ldots\hat{G}(E,\mathbf{k})V(\mathbf{k}-\mathbf{k}')\hat{G}(E,\mathbf{k}')\ldots$$
with
$$\ldots|\chi_\mathbf{k}\rangle
G(E,\mathbf{k})\langle\chi_{\mathbf{k}}|V(\mathbf{k}-\mathbf{k}')|\chi_\mathbf{k}'\rangle G(E,\mathbf{k}')
\langle\chi_{\mathbf{k}'}|\ldots $$
As a result, all the information about the E1-H1 structure as well as the chiral nature of particles
is now encoded in the angular dependence of the effective  amplitude of
scattering from a state $\mathbf k'$ into a state $\mathbf k:$
\begin{eqnarray}
\tilde{V}(\mathbf{k},\mathbf{k}')&=&\langle\chi_{\mathbf{k}}|V(\mathbf{k}-\mathbf{k}')|\chi_\mathbf{k}'\rangle
\nonumber
\\
&=&V(\mathbf{k}-\mathbf{k}')\frac{ 1+\eta\,\exp(\phi_{\mathbf{k}'}-\phi_\mathbf{k})}{ 1+\eta},
\label{tildeV}
\end{eqnarray}
where
\begin{equation}
\eta=\mu^2.
\label{eta-def}
\end{equation}
When $\eta=0$ the system is in the orthogonal symmetry class (the scattering amplitude has no angular dependence
due to Dirac factors), whereas the limit $\eta=1$ corresponds to the symplectic symmetry class with the disorder
scattering dressed by the ``Berry phase''.
The intermediate case corresponds to the unitary symmetry class, with a
competition between the Rashba-type and Zeeman-type
terms in the Hamiltonian killing the quantum interference.

We see that the problem is equivalent to a single-band problem with the Green's functions
\begin{equation}
G_{R,A}(E,\mathbf{k})=\frac{1}{E-E_\mathbf{k}\pm i\gamma/2}
\label{G+}
\end{equation}
and effective disorder potential dressed by ``Dirac factors'', Eq.~(\ref{tildeV}).
The quantum (total) scattering rate $\gamma$ entering the Green's function (\ref{G+}) as the imaginary
part of the self-energy is related to the disorder correlation function (\ref{corrW}) as follows:
\be
\gamma = \int_0^{2\pi} \frac{d\phi}{2\pi}~\gamma_D(\phi) =\gamma_0\frac{1+\eta^2}{(1+\eta)^2}, \label{gammaq}
\ee
where
\begin{eqnarray}
\label{gammaD}
\gamma_D(\phi_\mathbf k-\phi_{\mathbf k'})&=&\frac{2\pi}{\hbar} \int \frac{k dk' }{2\pi}
\langle |\tilde{V}(\mathbf{k},\mathbf{k}')|^2 \rangle \delta(E_\mathbf k-E_{\mathbf k'})
\nonumber\\
 &=&\gamma_0
 \frac{1+2\eta \cos(\phi_\mathbf k-\phi_{\mathbf k'}) +\eta^2}{(1+\eta)^2}
\end{eqnarray}
(here $\langle \cdots \rangle$ stands for disorder averaging),
\be
\gamma_0=\frac{2\pi \nu_F}{\hbar}  {W_0}
\ee
and
\be \nu_F=\frac{M}{2\pi\hbar^2 A^2}\frac{1+\eta}{1-\eta}\ee
is the density of states at the Fermi level (in a single cone per spin projection).

Analyzing the problem within the Drude-Boltzmann
approximation, it is easy to see that the rate $\gamma_D(\phi_\mathbf
k-\phi_{\mathbf k'})$ is the
rate of scattering from the momentum $\mathbf k'$ to the momentum  $\mathbf k.$
This function enters the collision integral of the kinetic equation and, as a
consequence, describes the vertex correlation function in the diffuson ladder.
(In the quasiclassical approximation, we can disregard the momentum $q$
transferred through disorder lines in these factors.) Though we consider the
short-range scattering potential, the function $\gamma_D(\phi_\mathbf
k-\phi_{\mathbf k'})$ turns out to be angle-dependent due to the ``dressing'' by
the spinor factor $\left| \left\langle \chi_\mathbf k | \chi_{\mathbf k'}
\right\rangle \right|^2.$  Hence, for the case of a massive Dirac cone, the
transport  scattering rate
\be
\gamma_{tr}= \int_0^{2\pi} \frac{d\phi}{2\pi}~\gamma_D(\phi)(1-\cos\phi)
=\gamma_0\frac{1+\eta^2-\eta}{(1+\eta)^2}
\ee
 differs from the total (quantum) rate $\gamma_q=\gamma:$
 \be
 \frac{\gamma}{\gamma_{tr}}=\frac{1+\eta^2}{1+\eta^2-\eta}.
 \label{gammatr}
 \ee

  \subsection{Kinetic equation for the Cooperon}
 \label{Coo}

It is well known that the Cooperon propagator obeys a kinetic equation.
\cite{AA,schmid,AAG}
The collision integral of this equation contains both incoming
and outgoing terms describing the scattering from a momentum $\mathbf k'$ into
a momentum  $\mathbf k.$
Importantly, the rates entering  these two terms  are different for the case of single massive cone.
The outgoing rate is determined by the rate $\gamma$ [which is the
angle-averaged function $\gamma_D(\phi)$] that
enters the single-particle Green function \eqref{G+}.
To find the incoming rate we notice that the disorder vertex lines in the Cooperon propagator
are also dressed by the Dirac spinor factors. Disregarding the momentum  transferred through disorder
lines in these factors, we find that  the  vertex line corresponding to the
scattering from  $\mathbf k'$ to   $\mathbf k$  is dressed by
 $$ \left\langle \chi_\mathbf k | \chi_{\mathbf k'}  \right\rangle \left\langle
\chi_\mathbf{ -k} | \chi_{\mathbf {-k'}}  \right\rangle.$$
 The corresponding rate $\gamma_C(\phi_\mathbf{k}-\phi_{\mathbf{k}'})$ is given
by Eq.~\eqref{gammaD} with  $\langle
|\tilde{V}(\mathbf{k},\mathbf{k}')|^2 \rangle$ replaced by $\langle
\tilde{V}(\mathbf{k},\mathbf{k}')\tilde{V}^*(-\mathbf{k'},-\mathbf{k})\rangle,$
 yielding:
 \begin{eqnarray}
\gamma_C(\phi_\mathbf{k}-\phi_{\mathbf{k}'})&=&\gamma_0 \frac{
1+2\eta~ e^{-i (\phi_{\mathbf{k}}-\phi_{\mathbf{k}'})}+\eta^2 e^{-2i (\phi_{\mathbf{k}}-\phi_{\mathbf{k}'})}}{(1+\eta)^2}.
\nonumber
\\
\label{Wc}
\end{eqnarray}
Let us make two comments which are of crucial importance for further
consideration.  First, we note that
\be
\int d\phi\gamma_C(\phi)\neq \int
d\phi\gamma_D(\phi),
\ee
which means that the collision integral in the Cooperon
channel does not conserve the particle number. This implies in turn that  the
Cooperon propagator has a finite decay rate even in the absence of the inelastic
 scattering. \cite{Tkachov11}
Another important property is an asymmetry of  $\gamma_C(\phi).$ Indeed, as
seen from Eq.~\eqref{Wc},
\be \gamma_C(\phi)=\gamma_C^*(-\phi)\neq\gamma_C(-\phi). \label{asimm}\ee

Once the projection on the upper band and the associated dressing
of the disorder correlators in the Cooperon ladders have been implemented, the
evaluation of the correction to the conductivity reduces to the solution of a
kinetic equation for the Cooperon propagator in an effective disorder. The
latter is
characterized by the correlation functions (\ref{Wc}) in the incoming part of
the collision integral and by (\ref{gammaD}) in the outgoing term.
The kinetic equation for the zero-frequency Cooperon
$C_{\mathbf{q}}(\phi,\phi_0)\equiv C(\mathbf{q},\omega=0;\phi,\phi_0)$
has the form:
\begin{align}
&\left[1/\tau_\phi + i \mathbf q \mathbf v_F  ) \right]C_{\mathbf{q}}(\phi,\phi_0)=\gamma \delta(\phi-\phi_0)
\nonumber
\\
&+\int \frac{d\phi'}{2\pi} [\gamma_C(\phi-\phi') C_{\mathbf{q}}(\phi',\phi_0)-\gamma_D(\phi-\phi') C_{\mathbf{q}}(\phi,\phi_0)].
\label{Cooperon}
\end{align}
Here, $1/\tau_\phi$ is the phase-breaking rate,  $\mathbf v_F=v_F(\cos \phi, \sin \phi ),$ and
\be
v_F=\frac{2\sqrt{\eta}}{1+\eta}\, A
\ee
is the Fermi  velocity at the Fermi energy
\be
E_F=M\frac{1+\eta}{1-\eta}.
\label{EF}
\ee
 The Fermi wave vector is given by
 \be
 \label{kF}
 k_F=\frac{2M}{\hbar A} \frac{\sqrt{\eta}}{1-\eta}.
 \ee
 Diagrammatically, Eq.~(\ref{Cooperon}) corresponds to a Cooperon impurity
ladder with four Green's functions at the ends.

 Introducing dimensionless variables
\be
\Gamma= 1/\gamma \tau_\phi,\quad \mathbf Q= \mathbf q l,\ee
where
\be l=v_F/\gamma=\frac{2 \hbar^3 A^3}{M W_0} \frac{\sqrt{\eta}(1-\eta)}{1+\eta^2}
\ee
is the mean free path, we rewrite Eq.~\eqref{Cooperon} as follows:
\begin{align}
&(1 +  \Gamma+ i \mathbf Q \mathbf n ) C_{\mathbf{Q}}(\phi,\phi_0)
= \delta(\phi-\phi_0)\nonumber
\\
&+\int \frac{d\phi'}{2\pi} \frac{[1+\eta e^{-i(\phi-\phi')}]^2}{1+\eta^2} C_{\mathbf{Q}}(\phi',\phi_0)
,
\label{Cooperon1}
\end{align}
where $\mathbf n=(\cos \phi, \sin \phi ).$
As seen from Eq.~\eqref{Cooperon1}, the incoming term  of the collision integral contains only
three angular harmonics: $0,-1,-2.$  This allows us to present the solution of Eq.~\eqref{Cooperon1} in the following form:
\begin{align}
 &C_{\mathbf{Q}}(\phi,\phi_0)
 =  \frac{1}{1 +  \Gamma+ i \mathbf Q \mathbf n}\nonumber
\\
&\times [C_0+e^{i(\phi_\mathbf{Q}-\phi)}C_{-1}+e^{2i(\phi_\mathbf{Q}-\phi)}C_{-2}+\delta(\phi-\phi_0)],
\label{Cooperon0-1-2}
\end{align}
where
\bee
C_{0}&=& \frac{1}{1+\eta^2}\int \frac{d\phi}{2\pi} C_{\mathbf{Q}}(\phi,\phi_0),
\label{C0}\\
C_{-1}&=& \frac{2\eta}{1+\eta^2}\int \frac{d\phi}{2\pi}
C_{\mathbf{Q}}(\phi,\phi_0)e^{i(\phi-\phi_\mathbf{Q})},\label{C-1}\\
C_{-2}&=& \frac{\eta^2}{1+\eta^2}\int \frac{d\phi}{2\pi}
C_{\mathbf{Q}}(\phi,\phi_0)e^{2i(\phi-\phi_\mathbf{Q})},\label{C-2}
\eee
and $\phi_\mathbf{Q}$ is the polar angle of vector $\mathbf{Q}.$ Substituting Eq.~\eqref{Cooperon0-1-2}
into Eqs.~\eqref{C0}, \eqref{C-1} and \eqref{C-2},
we find a system of coupled equations  for $C_0,C_{-1},$ and $C_{-2}$ which can be written in the matrix form
\be \label{MC}
\hat M  \left[\begin{array}{cc} C_0 \\C_{-1} \\ C_{-2}\end{array}\right]
=\frac{1}{2\pi(1 +  \Gamma+ i \mathbf Q \mathbf n_0)}
\left[\begin{array}{cc} 1 \\e^{i(\phi_0-\phi_\mathbf{Q})} \\ e^{2i(\phi_0-\phi_\mathbf{Q})} \end{array}\right].
\ee
Here
\begin{equation}\label{M}
\hat{M}=\left[\begin{array}{cccc} 1+\eta^2-P_0 & - P_1 & - P_2\\
-P_1 & \dfrac{1+\eta^2}{2\eta} - P_0 & - P_1 \\
-P_2 & - P_1 & \dfrac{1+\eta^2}{\eta^2} - P_0
\end{array}\right],
\end{equation}
\begin{align}
P_n&=\int \frac{d\phi}{2\pi} \frac{e^{-i n \phi}}{1 + \Gamma + i Q \cos\phi }
\nonumber
\\
 &=(-i)^{|n|}P_0 \left[\frac{1-P_0(1+\Gamma)}{1+P_0(1+\Gamma)} \right]^{|n|/2},
\label{Pn}
\end{align}
and
\be
P_0=\frac{1}{\sqrt{(1+\Gamma)^2+Q^2}}.
\ee
From Eqs.~\eqref{Cooperon0-1-2}, \eqref{MC}, and \eqref{M} we find
\begin{eqnarray}
&&C_{\mathbf{Q}}(\phi,\phi_0)
=\frac{\delta(\phi-\phi_0)}{1 +  \Gamma+ i \mathbf Q \mathbf n}
\nonumber \\
&&
+\frac{1}{2\pi(1 +  \Gamma+ i \mathbf Q \mathbf n)
 (1 +  \Gamma+ i \mathbf Q \mathbf n_0)}
 \nonumber \\
 && \times
 \left[\begin{array}{cc} 1 \\e^{i(\phi_\mathbf{Q}-\phi)} \\ e^{2i(\phi_\mathbf{Q}-\phi)} \end{array}\right]^T \hat M^{-1}
 \left[\begin{array}{cc} 1 \\e^{i(\phi_0-\phi_\mathbf{Q})} \\ e^{2i(\phi_0-\phi_\mathbf{Q})} \end{array}\right],
  \label{Cooperon2}
\end{eqnarray}
where $\mathbf n=(\cos \phi_0, \sin \phi_0 ).$  The Fourier transform of the
Cooperon propagator gives  the quasiprobability~\cite{schmid} (per unit area)
for an electron starting with a momentum  direction $\mathbf n_0$ from an
initial point  $\mathbf r_0$    to arrive at a point $\mathbf r$
with a momentum direction $\mathbf n:$
\be
C(\phi,\phi_0,\mathbf r-\mathbf r_0)=\frac{1}{l^2}\int\frac{d^2\mathbf Q}{(2\pi)^2}e^{i\mathbf Q(\mathbf r- \mathbf r_0)/l}C_\mathbf Q (\phi,\phi_0).
\label{Cooperon-coord}
\ee
In particular, the conductivity can be expressed in terms of this probability taken at $\mathbf r-\mathbf r_0=0$  (return probability):
 \be
  W(\phi-\phi_0)=C(\phi,\phi_0,0).
  \ee

The first term in the r.h.s. of  Eq.~\eqref{Cooperon2} describes the ballistic
motion (no collisions). The second term can be expanded (by expanding the
matrix $M^{-1}$) in series over functions $P_n.$ Such an expansion is, in fact,
an expansion of the Cooperon propagator over the number $N$ of collisions (the
zeroth term in this expansion corresponds to $N=1$).\cite{nonback} Since the
term with $N=1$ does not contribute to the interference-induced
magnetoresistance, we can exclude it from the summation in the interference
correction and regard this contribution as a part of the Drude
conductivity.\cite{comment}  Indeed, after a substitution into $W(\phi,\phi_0)$
we see that this term describes a return to the initial point after a single
scattering, so that the corresponding trajectory does not cover any area and,
consequently, is not affected by the magnetic field. Neglecting both
the ballistic ($N=0$) and the $N=1$ terms in the Cooperon propagator, we find
\begin{eqnarray}
\label{Cooperon3}
&& C_{\mathbf{Q}}(\phi,\phi_0)
=\frac{1}{2\pi(1 +  \Gamma+ i \mathbf Q \mathbf n)
 (1 +  \Gamma+ i \mathbf Q \mathbf n_0)}
 \nonumber \\
 &&
 \times \left[\begin{array}{cc} 1
\\e^{i(\phi_\mathbf{Q}-\phi)} \\ e^{2i(\phi_\mathbf{Q}-\phi)}
\end{array}\right]^T
 \left (\hat  M^{-1}-\hat M^{-1}_{Q=\infty} \right)\left[\begin{array}{cc} 1
\\e^{i(\phi_0-\phi_\mathbf{Q})} \\ e^{2i(\phi_0-\phi_\mathbf{Q})}
\end{array}\right]. \nonumber \\ &&
\end{eqnarray}
Here we took into account that $P_n \to 0$ for $Q\to\infty.$  Let us now find the return probability.
To this end, we  make expansions
\begin{align}
\frac{1}{1 +  \Gamma+ i \mathbf Q \mathbf n}&=\sum_{n=-\infty}^{\infty}P_n
e^{in(\phi-\phi_\mathbf Q)},
\\
 \frac{1}{1 +  \Gamma+ i \mathbf Q \mathbf n_0}&=\sum_{m=-\infty}^{\infty}P_m e^{-im(\phi_0-\phi_\mathbf Q)}
\end{align}
 in Eq.~\eqref{Cooperon3}, substitute the obtained equation into
Eq.~\eqref{Cooperon-coord}, take $\mathbf r=\mathbf r_0,$   and average over
$\phi_\mathbf Q,$.  We arrive then to the following equation
 \be \label{W}
 W(\phi)=\frac{1}{2 \pi l^2}\sum_{n=-\infty}^{\infty} w_n e^{i(n-1)\phi},
 \ee
 where
 \begin{eqnarray}
 w_n=\int \frac{d^2\mathbf Q}{(2\pi)^2} \left[\begin{array}{cc} P_{n-1} \\ P_{n} \\ P_{n+1} \end{array}\right]^T
 \left (\hat  M^{-1}-\hat M^{-1}_{Q=\infty} \right)\left[\begin{array}{cc} P_{n-1} \\ P_{n} \\ P_{n+1} \end{array}\right].
 \nonumber
 \\
 \label{wn}
 \end{eqnarray}

On a technical level, the logarithmic divergency
specific for WL and WAL conductivity corrections comes from a singular behavior
of the matrix $\hat M^{-1}$ at $Q\to 0.$ Before analyzing the solution in the
full generality, let us consider the limiting case
$Q=0$, $\Gamma=0$. In this case $P_0=1$, $P_1=P_2=0$, and we find:
 \begin{equation}
\hat M^{-1}=\begin{pmatrix} \dfrac{1}{\eta^2} & 0 & 0\\
0 & \dfrac{2\eta}{(1-\eta)^2} & 0 \\
0 & 0 & \eta^2
\end{pmatrix}.
\label{M-1}
\end{equation}
Then in the limit $\eta\to 0$ (orthogonal class) the singular mode is $C_0$ [see Eq.~\eqref{Cooperon0-1-2}] and $C_\mathbf Q(\phi,\phi_0) \propto 1/\eta^2$,
while in the limit $\eta\to 1$ (symplectic class) the singular mode is $C_{-1}$ and $C_\mathbf Q(\phi,\phi_0) \propto e^{-i(\phi-\phi_0)}/(1-\eta)^2$.

\subsection{Correction to the conductivity}
\label{Corr}

 The quantum correction to the conductivity is given by~\cite{GKO14}
\be \label{sigma}
\delta \sigma=
-\frac{e^2}{ \hbar}l_{tr}^2~\int \frac{d\phi}{2\pi}~\frac{\gamma_C(\pi -\phi)}{\gamma} W(\phi)(1+\cos\phi).
\ee
Here in the ``Hikami-box'' factor $1+\cos\phi$ the unity comes from the conventional Cooperon diagram describing the backscattering
contribution, while $\cos\phi$ arises from a Cooperon covered by an impurity line (nonbackscattering term~\cite{nonback}).
Using Eqs.~\eqref{gammaq}, \eqref{gammatr}, and \eqref{Wc}, we obtain
\begin{align}
\delta \sigma&=
-\frac{e^2}{ \hbar} \frac{(1+\eta^2)l^2}{(1+\eta^2-\eta)^2}
\nonumber
\\
&\times
\int \frac{d\phi}{2\pi} (1-2\eta e^{i\phi}+\eta^2 e^{2i\phi})  W(\phi)(1+\cos\phi),
\label{sigma1}
\end{align}
and, finally, with the use of Eq.~\eqref{W}, arrive at
\begin{eqnarray}
 \label{sigma2}
\delta \sigma&=&-\frac{e^2}{2\pi \hbar}
\frac{1+\eta^2}{(1+\eta^2-\eta)^2}\left[(1-\eta)w_1+\frac{1+\eta^2-4\eta}{2}
w_0 \right. \nonumber \\
&+& \left.\frac{w_2}{2}+(\eta^2-\eta)w_{-1}+\frac{\eta^2}{2} w_{-2}
\right],
\end{eqnarray}
where $w_n$ are given by Eq.~\eqref{wn}.

%%%%%%%%%%%%%%%%%%%%%%%%%%%%%%%%%%%%%%%%%%%%%%%%%%%%%%%
\begin{figure*}
\includegraphics[width=0.45\textwidth]{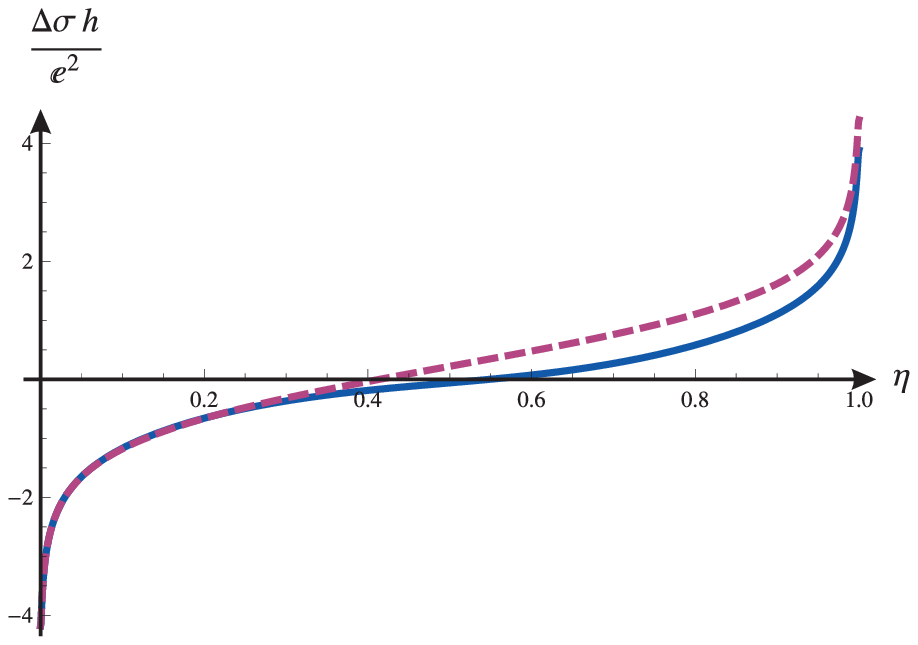}
\includegraphics[width=0.43\textwidth]{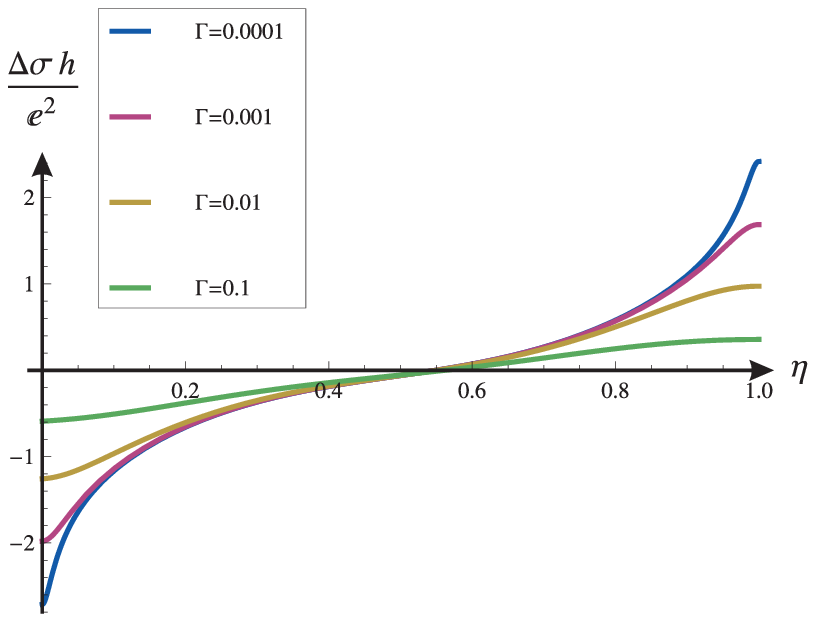}
\caption{\emph{Left panel:} Conductivity correction in the absence of block
mixing, Eq.~\eqref{sigma2},
as a function of $\eta$ for infinite dephasing time (solid line); sum of the
logarithmic WL and WAL asymptotics,
Eqs.~\eqref{logeta0} and \eqref{logeta1} (dashed line).
\emph{Right panel:} Conductivity correction for different values of
dimensionless dephasing rate $\Gamma$.
}
\label{FigI}
\end{figure*}

\begin{figure*}
\centerline{\includegraphics[width=0.45\textwidth]{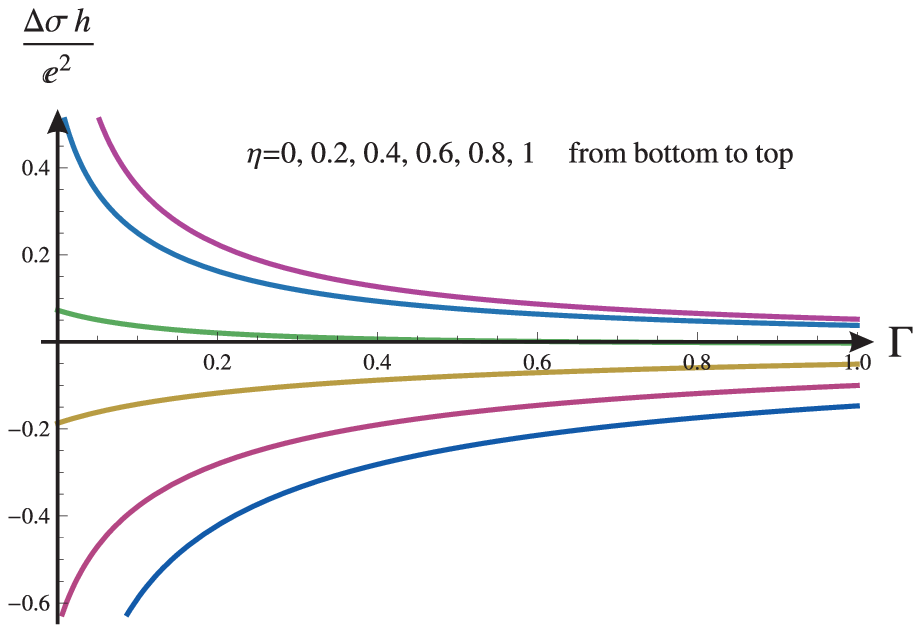}
\includegraphics[width=0.45\textwidth]{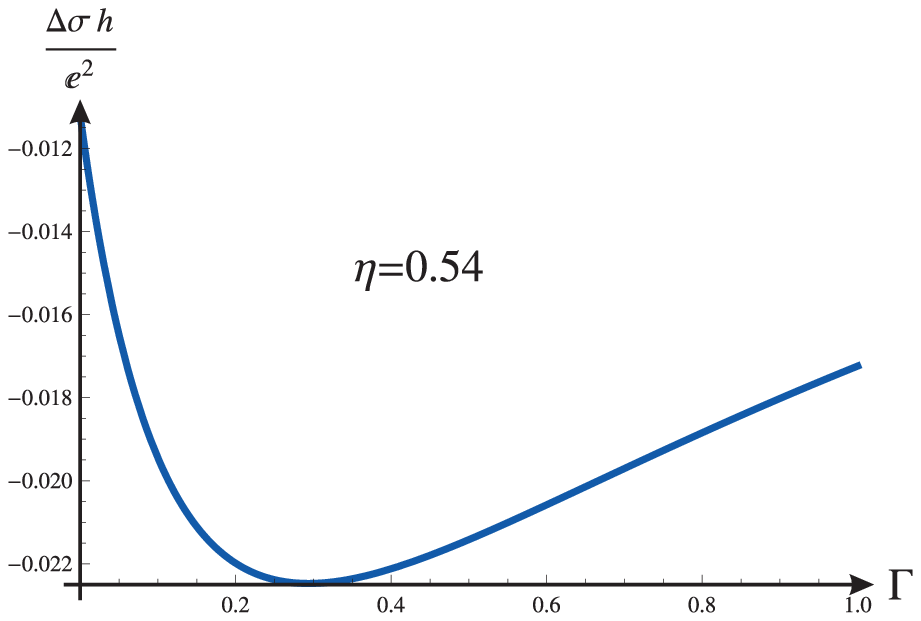}}
\caption{
\emph{Left panel:} Conductivity correction in the absence of block mixing as a
function of the dephasing rate for different values of $\eta$.
\emph{Right panel:} Conductivity correction as a function of dephasing rate for
$\eta=0.54$.}
\label{FigII}
\end{figure*}
%%%%%%%%%%%%%%%%%%%%%%%%%%%%%%%%%%%%%%%%%%%%%%%%%%%%%%%

As discussed above, for $\eta=0$ and $\eta=1$ one of the modes becomes singular
(corresponding to $C_0$ and $C_{-1},$ respectively). Keeping the singular modes only,
one can easily obtain  the return  probability and the  conductivity  in vicinities  of the points
$\eta=0$ and $\eta=1.$   For $\eta\to 0$
we find
\be W(\phi)\approx \frac{1}{2\pi l^2} \int\frac{d^2\mathbf Q}{(2\pi)^2}\frac{P_0^3}{1+\eta^2-P_0}
\ee
and
\be \delta \sigma\approx-\frac{e^2}{4\pi^2\hbar}\ln\left(\frac{1}{\eta^2+\Gamma}\right).\label{logeta0}\ee
For $\eta\to 1$
we find
\be W(\phi)\approx \frac{e^{-i\phi}}{2\pi l^2}  \int\frac{d^2\mathbf Q}{(2\pi)^2}\frac{P_0^3}{1+(1-\eta)^2/2\eta-P_0-2P_1^2},
\label{Wlim1}
\ee
and
\be \delta \sigma\approx\frac{e^2}{4\pi^2\hbar}\ln\left[\frac{1}{(1-\eta)^2/2+\Gamma}\right].\label{logeta1}\ee
According to Eqs.~(\ref{logeta0}) and (\ref{logeta1}), the symmetry breaking
rates $1/\tau_A$ and $1/\tau_m$ introduced in Sec.~\ref{s3} are equal to
\be
\label{tauA}
1/\tau_A = \eta^2 \gamma \qquad (\eta \ll 1)
\ee
and
\be
\label{taum}
1/\tau_m = (1-\eta)^2 \gamma/2 \qquad (1-\eta \ll 1),
\ee
respectively.

Using Eqs. (\ref{eta-def}) and (\ref{mu-def}), the result (\ref{taum}) for the symmetry-breaking rate $1/\tau_m$
in the range of almost linear spectrum ($1-\eta\ll 1$) agrees with the result of
Ref. \onlinecite{Tkachov11}, where this scale was first identified.
In the opposite limit $\eta\to 0$ (which was not analyzed in Ref. \onlinecite{Tkachov11}),
the result (\ref{tauA}) of the microscopic calculation confirms the estimate of Ref. \onlinecite{OGM12}.

The leading logarithmic contributions (\ref{logeta0}) and (\ref{logeta1}) are
exactly the WL and WAL corrections found in Sec.~\ref{s3} from the symmetry
analysis for the regimes 2O and 2Sp, respectively. We are now in a position to
evaluate corrections to these results. Setting for simplicity $\Gamma=0$ (no
dephasing), we get from Eqs.~(\ref{sigma2}) and (\ref{wn})
\begin{widetext}
\be
\delta \sigma \approx \frac{e^2}{4\pi^2 \hbar}
\begin{cases}
\displaystyle -\ln\frac{1}{\eta^2} +\ln2
+(4\ln 2 -1) \eta +   \frac{5}{2}\eta^2 \ln\frac{1}{\eta^2}  + O(\eta^2) ,&\qquad
{\rm for}~ \eta\to 0,\\[0.3cm]
\displaystyle \ln\frac{1}{(1-\eta)^2}
+3(\ln2-1) - (1-\eta)
-\frac{7}{2}(1-\eta)^2\ln\frac{1}{(1-\eta)^2}
+O[(1-\eta)^2],&\qquad  {\rm for}~ \eta\to 1.
\end{cases}
\label{asympt}
\ee
\end{widetext}
The regime $\eta\to 1$ was considered in Ref.~\onlinecite{Tkachov11}.
However, our results for subleading terms in this regime differ from those obtained there.
This difference is due
to the fact that Ref.~\onlinecite{Tkachov11} only considered contributions of
small momenta $ql\ll 1$. While this is sufficient to get the universal WL and
WAL terms (\ref{logeta0}) and (\ref{logeta1}), evaluation of the corrections
requires taking into account the ``ballistic'' momenta $ql \gtrsim 1$.

We note that the last term, $\mathcal{O}[(1-\eta)^2 \ln(1-\eta)],$ in the asymptotics for $\eta\to 1$ in Eq.~(\ref{asympt})
is, in fact, determined by momenta $ql\ll 1$ and was identified within the diffusive approximation
of Ref.~ \onlinecite{Tkachov11}. However, the numerical coefficient in front of this term in Ref. \onlinecite{Tkachov11},
which in our notation would take the form $-(19/8)(1-\eta)^2 \ln[1/(1-\eta)^2]$, differs from
the prefactor $-7/2$ in the corresponding term in our Eq.~(\ref{asympt}).
The difference stems from setting $q=0$ everywhere (except for the term $Dq^2$ in the denominators of the Cooperon propagators)
in the calculation of Ref. \onlinecite{Tkachov11}. It is worth emphasizing that the ballistic terms
$\mathcal{O}(1)$ and $\mathcal{O}(1-\eta)$ neglected in Ref. \onlinecite{Tkachov11}
give a much larger contribution in the limit $\eta\to 1$ than this ``diffusive-like'' term.

The results for arbitrary values of $\eta$ are shown in Figs. \ref{FigI} and
\ref{FigII}. In the  left panel  of Fig.~\ref{FigI} we plotted
the conductivity correction calculated with the use of Eqs.~\eqref{wn} and
\eqref{sigma2} (solid line) in the absence of dephasing ($\Gamma=0$). The  sign
of the correction changes with increasing $\eta,$ so that the system undergoes
a crossover from the orthogonal to symplectic ensemble as  expected.
In the same panel we plotted by dashed line the sum of two asymptotes, given by
Eqs.~\eqref{logeta0} and \eqref{logeta1}, respectively.
A deviation of the exact result from the interpolating formula
is mostly due to the nonsingular mode related to the eigenvalue $\eta^2$ in
matrix ${\hat M}^{-1}$ given by Eq.~\eqref{M-1}. As a result, the  difference
between the solid and dashed lines vanishes in the limit $\eta\to 0$, while
yielding a shift $\sim 1$ near $\eta\to 1$.
The right panel of Fig.~\ref{FigI} illustrates the crossover at different
dephasing rates.  In  the left panel of Fig.~\ref{FigII} we plotted
the conductivity correction as a function of dephasing rate for different
$\eta.$ The most interesting feature of these curves is the
nonmonotonous  dependence  of conductivity on $\Gamma$ for $\eta$ close to the
point separating localization and delocalization behavior. This feature is
emphasized in the right panel of Fig.~\ref{FigII}, where dependence of
conductivity on dephasing rate is plotted for $\eta=0.54.$  We expect
that the conductivity in the 2U symmetry regime (energy $E_3$ in
Fig.~\ref{Fig1}) shows an analogous non-monotonous dependence on the magnetic
field due to ballistic effects.

\subsection{Block mixing}
\label{Mix}
Let us now briefly analyze the effect of a weak block mixing.
As discussed above,
on largest scales the mixing leads to a single copy of the symplectic class
(1Sp).  Below we assume the simplest form of the mixing potential:
\begin{equation}\label{Vmix}
\hat{V}=V(\mathbf r)\left[\begin{array}{cccc} 1 & 0 & 0& -\Delta\\
0 & 1& \Delta & 0\\
0 & \Delta^* & 1 &0\\ -\Delta^* &0&0&
\end{array}\right],
\end{equation}
where $V(\mathbf r)$ is a short-range potential with the correlation function given by Eq.~\eqref{corrW}
and $\Delta$ is a parameter responsible for the block mixing.
We also assume that $\Delta $ is small and real: $\Delta \ll 1,~\Delta=\Delta^*.$
The potential Eq.~\eqref{Vmix} obeys the symmetry: $\hat V=\hat U \hat V^* \hat U^{-1},$ where
\begin{equation}\label{U}
\hat{U}=\left[\begin{array}{cccc} 0 & 0 & -i& 0\\
0 & 0& 0 & -i\\
i & 0& 0 &0\\ 0 &i&0&0
\end{array}\right],~~\hat U \hat U^*=-1.
\end{equation}

The result of the microscopic calculation reads \cite{GKO14}:
\begin{eqnarray}
\delta \sigma&=&\frac{e^2}{4\pi^2\hbar}
\left\{
-\ln \left[\Gamma+\frac{(1-\eta)^2}{2}\right]
\right.
\nonumber \\
&-&
\ln \left[\Gamma + \frac{(1-\eta)^2}{2}+2\Delta^2\right]
  + 2\ln\big[\Gamma+\eta^2 +2\eta\Delta^2 \big]
\nonumber \\
 &+& \left. \ln \left[\Gamma +\frac{4\Delta^2\eta}{1+\eta^2}\right] - \ln \Gamma
\right\}. \label{bl-mix}
\end{eqnarray}
We stress that this equation was obtained in the diffusion approximation under
the assumption that
dimensionless gaps of diffusive modes are small. Hence, though expressions for
the last two logarithms in this equation are exact,
the explicit form  of  the first two logarithms is exact only in a vicinity of
the point $\eta =1,$ while the third logarithm
is exact near the point $\eta= 0.$
In this sense,  Eq.~\eqref{bl-mix} interpolates between  the two limits ($\eta=0$ and $\eta=1$) similarly to
Eq.~(\ref{interpol}) (cf. Fig. \ref{FigI}).
Away from the points $\eta=0$ and $\eta=1$ the block mixing in the first three terms of Eq.~\eqref{bl-mix}
can be neglected, so that one can use Eq.~\eqref{sigma2} with Eq.~(\ref{wn}) to describe these terms in the crossover range of $\eta$
beyond the diffusive approximation.
It is also worth noting that the gap $4\Delta^2\eta/(1+\eta^2)$ in the fourth logarithm is nonzero only when both
$\Delta$ and $\eta$ are nonzero, see discussion above Eq.~\eqref{OFull}.

Introducing the symmetry breaking times $\tau_m$, $\tau_A$, and $\tau_\Delta$ according to
\begin{eqnarray}
\frac{\tau}{\tau_m}&=&\frac{(1-\eta)^2}{2},
\\
\frac{\tau}{\tau_A}&=&\eta^2+2\eta \Delta^2,
\\
\frac{\tau}{\tau_\Delta}&=&\frac{4\Delta^2(\eta+2\Delta^2)}{1+\eta^2},
\end{eqnarray}
we can rewrite Eq.~(\ref{bl-mix}) in a form combining (for $1/\tau_{SO}=0$)
Eqs.~(\ref{SpFull}) and (\ref{OFull}):
\begin{eqnarray}
\delta\sigma
   &\simeq&
   \frac{e^2}{2 \pi h} \Bigg[
   -\ln\left(
       \frac{\tau}{\tau_\phi}+\frac{\tau}{\tau_m}
     \right)
     -\ln\left(
       \frac{\tau}{\tau_\phi}+\frac{\tau}{\tau_m}+\frac{\tau}{\tau_\Delta}
     \right) \nonumber \\
&+&  2 \ln\left(
       \frac{\tau}{\tau_\phi}+\frac{\tau}{\tau_A}
     \right)
     +\ln\left( \frac{\tau}{\tau_\phi}
     +\frac{\tau}{\tau_A/2 +\tau_\Delta}\right)
     \nonumber \\
 &-&  \ln \frac{\tau}{\tau_\phi} \Bigg].
     \label{Delta-eta-Full}
\end{eqnarray}
This expression correctly reproduces all singular interference corrections in
the whole range $0 \leq \eta \leq 1$ provided $\Delta \ll 1$.
Note also that for $\eta\ll \Delta^2$, Eq.~\eqref{Delta-eta-Full} reduces to the
conventional expression for the symmetry pattern 2O $\to$ 1Sp with anisotropic
spin relaxation (different for in-plane and out-of-plane components)
described by $\tau_{||}=\tau_A$ and $\tau_\perp \simeq \tau_A/2=\tau_{||}/2$.
In a similar way, one can also include the block mixing described by $1/\tau_{SO}$.

\section{Weak antilocalization in a quasi-one-dimensional strip}
\label{s5}

Motivated by recent experiment of Ref. \onlinecite{bruene},
we will discuss in this section the magnetoresistance of a quasi-one-dimensional
HgTe structure.
The magnetotransport measurements in Ref. \onlinecite{bruene} were performed
on a 2D quantum wells patterned in narrow 1D strips. While the width of the
HgTe quantum well is of the order of few nanometers, its lateral width
$W$ can be less than hundred nanometers. In such a restricted
quasi-1D geometry, boundary conditions play a crucial role for electron
transport. Below we analyze possible symmetries of the boundary
conditions and ballistic transport effects arising in quasi-1D samples.

\subsection{Symmetry breaking mechanism due to boundary scattering}
\label{s2.3}

We start by assuming that the boundary does preserve the $s_z$-symmetry
and the Hamiltonian splits into two independent blocks $h(\mathbf{k})$ and
$h^*(-\mathbf{k})$ related by the physical TR symmetry.
It was demonstrated in Ref. \onlinecite{OGM12} that in this case the block-wise
TR-symmetry of $h(\mathbf{k})$ is necessarily broken by boundary conditions.

It is well known that a massless Dirac electron cannot be confined by any potential
profile due to the Klein tunneling effect. The only way to introduce
boundary conditions for a single flavor of Dirac fermions (without block
mixing)
is to open a gap at the boundary by introducing a large mass in the
Hamiltonian. This boundary mass term violates the block-wise symplectic TR symmetry.
On a technical level, each boundary scattering event in a Dirac system introduces
an additional phase shift in the electron wave function. These scattering phases
combine in such a way that the amplitudes of two mutually time-reversed
trajectories pick up a relative phase difference $\pi$ for each boundary scattering.
Thus the contribution of a trajectory with $N$ boundary
scattering events to the conductivity correction contains a factor $(-1)^N$. After the summation
over all paths with arbitrary $N$, this alternating sign factor kills the singular correction
to the conductivity.

More general boundary conditions are possible if the massive BHZ Hamiltonian
includes higher-order (quadratic) terms in the $k$ expansion. In particular, the
hard-wall boundary conditions, $\psi = 0$, are widely used in the literature in
this case. The TR symmetry breaking mechanism discussed above is however still
effective in this limit provided $E_F\gg |m(k_F)|$. Quadratic terms become
important only very close to the boundary, while at a longer distance the dynamics
is well described by the linearized Hamiltonian. The hard-wall model
effectively
reduces to the same infinite mass boundary condition for the problem at these
longer scales. An explicit computation of the corresponding scattering states
can be found in Ref.\ \onlinecite{OGM12}.

We arrive at the conclusion that the block-wise symplectic TR-symmetry is
inevitably broken by the boundary conditions.
In a diffusive quasi-1D strip, the edge-induced symmetry-breaking rate reads
\begin{equation}
 \frac{1}{\tau_\text{edge}}
  \sim \frac{D}{W^2},
\end{equation}
where $W$ is the strip width. The time $\tau_\text{edge}$ is given by the
average time
for a particle to propagate diffusively between the edges and ``to become
aware'' of the violation
of the block-wise TR symmetry by the boundaries.
When the mean free path is longer than the width of the
quasi-1D sample, the corresponding symmetry breaking time is given by the
typical flight time for ballistic propagation between the two boundaries.
The total rate of breaking the block-wise symplectic TR symmetry
at $\eta$ close to unity is thus given by $1/\tau_m+1/\tau_\text{edge}$.

In the opposite limit $\eta\to 0$, when the block-wise TR symmetry
is of the orthogonal type, the vanishing of the wave function at the boundary
does not lead to the TR symmetry breaking. Indeed, conventional
``Schr\"odinger'' particles
can be confined by a scalar potential.

Above we have discussed boundary conditions that preserve the $s_z$-symmetry.
The spin symmetry, however, can be broken at boundaries of a 2D sample if the
edges violate the reflection symmetry in $z$-direction. Such edges can be modeled by short-range
impurities located near the boundaries. This situation is quite likely for realistic
samples. In this case, the only remaining symmetry respected by the edges
is the physical symplectic TR symmetry.

\subsection{Interference corrections and magnetoresistance in a
  quasi-1D system}
\label{s:1D}

Let us now analyze the interference effects in a quasi-1D geometry, i.e. in a
strip
of width $W$. We assume a single copy of symplectic class since the block-wise
TR symmetry is broken by the boundaries, so that only
the patterns 1Sp or 2U $\to$ 1Sp
are realized due to the block mixing with no room for 2Sp regime. The magnetoresistance
depends on the relations between the width $W$, 2D mean free path $l$,
dephasing length $L_\phi$, and magnetic length $L_H$. We start our analysis
with the case of sufficiently weak dephasing $L_\phi \gg W$. In the opposite
limit, when the dephasing length is shorter than the width of the strip, the
WAL correction is given by 2D formulas. For $W \gg L_\phi \gg l$ the diffusive
results apply, whereas for $W\gg L_\phi \alt l$ the WAL correction is described
by nonuniversal ballistic 2D formulas; the magnitude of the correction is small
in this case.

For sufficiently weak magnetic fields, the characteristic length scale at which
the interference is suppressed is determined by the condition
\begin{equation}
L_0 \sim L_H^2/W.
\end{equation}
For such lengths, the area $L_0 W$ is pierced by one magnetic flux. We assume
$L_0 \gg l$, thus the motion along the strip is diffusive within $L_0$. This
implies sufficiently weak magnetic field, $L_H \gg \sqrt{W l}$.

When the width of the quasi-1D strip is much larger than the 2D mean free path,
$W \gg l$, the correction to the conductance of the strip of length $L$ due to
WAL reads:\cite{AltshulerAronov81,Beenakker,AA,AAG}
\begin{equation}
 \delta G
  = \frac{e^2}{h} \frac{1}{L} \left(
      \frac{W^2}{3L_H^4} + \frac{1}{L_\phi^2}
    \right)^{-1/2}.
    \label{diffuse}
\end{equation}
Note that this expression assumes a phenomenological ($B$-independent) dephasing
length; a more accurate formula including Airy function can be found in Refs.
\onlinecite{AAK,AA,AAG}. The difference between the two expressions is of the order of
few percent, so that we use the formula (\ref{diffuse}). The measured
correction to the longitudinal conductivity of the 2D array of strips is given
by
\begin{equation}
\delta\sigma=\frac{L}{W}\delta G,
\end{equation}
neglecting the separation between the strips.

For narrow strips, $W\alt l$, the transverse motion is ballistic and the
resulting correction is modified,
\begin{equation}
 \delta \sigma
  = \frac{e^2}{h} \left(
      \frac{W^4}{3L_H^4} f(W/l) + \frac{W^2}{L_\phi^2}
    \right)^{-1/2}. \label{ballistic}
\end{equation}
The function $f(W/l)$ incorporates information about boundary conditions.
For diffuse boundaries in the quasi-2D geometry various limiting cases of this
function were obtained in Ref.\ \onlinecite{DK}. In a later paper Ref.\
\onlinecite{Beenakker}, this function was computed numerically for both
quasi-1D and quasi-2D samples with both mirror and diffuse boundaries. In the
limiting case $W \ll l$, analytical expression was also obtained for diffuse
boundary conditions.

\begin{figure}
\centerline{\includegraphics[width=0.43\textwidth]{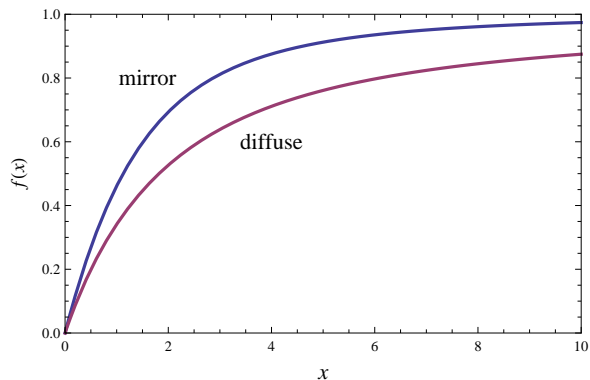}}
\caption{Functions $f_\text{mirror}(x)$ and $f_\text{diffuse}(x)$ defined in
Eqs. (\ref{fx}).}
\label{Fig1D}
\end{figure}

We represent the quasi-1D results in a closed integral form applicable for
all values of $x = W/l$:
\begin{widetext}
\begin{subequations}
\label{fx}
\begin{align}
 f_\text{mirror}(x)
  &= 1 - \frac{3}{x^2} + \frac{96}{\pi x^3} \int ds\, s^3 \sqrt{1 - s^2}
      \tanh\left(\frac{x}{2s}\right)
  \approx \begin{cases}
    1, & x \gg 1, \\
    \dfrac{186 \zeta(5)}{\pi^5}\; x \approx 0.63\; x, & x \ll 1,
  \end{cases} \\
 f_\text{diffuse}(x)
  &= 1 - \frac{4}{\pi x} + \frac{32}{5 \pi x^3} - \frac{12}{\pi x} \int
ds\, s \sqrt{1 - s^2} e^{-x/s} \left(1 + \frac{4s}{x} + \frac{4s^2}{x^2}\right)
  \approx \begin{cases}
    1, & x \gg 1, \\
    \dfrac{3x}{2\pi}, & x \ll 1.
  \end{cases}
\end{align}
\end{subequations}
\end{widetext}
These functions are shown in Fig. \ref{Fig1D}. For $x\equiv W/l \gg 1$, the
functions
$f(x)$ approach the universal value corresponding to the diffusive result Eq.\
(\ref{diffuse}). Ballistic effects partly suppress WAL due to cancellation of
magnetic flux piercing purely ballistic trajectories.

Remarkably, the magnitude of the magnetoconductivity peak in quasi-1D geometry
is determined only by the ratio of the dephasing length $L_\phi$ and the strip
width $W$,
\begin{equation}
\delta\sigma(B=0)=\frac{e^2}{h} \frac{L_\phi}{W}.
\label{peak}
\end{equation}
Details of transverse ballistic motion, encoded in the function $f(W/l)$, are
relevant only for the width of the peak.

\begin{figure}
\centerline{\includegraphics[width=0.43\textwidth]{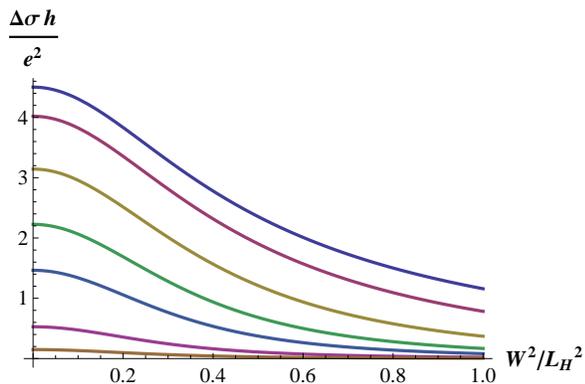}}
\caption{Conductivity correction in a quasi-1D strip of HgTe,
Eq.~(\ref{1Dbm}), for $W/l\gg1$, $W/L_\phi=0.2,$
and different values of the block-mixing length: $W/L_{SO}=2,\,1,\,0.5,\,0.3,\,0.2,\,0.1,\,0.05$
from top to bottom.}
\label{Fig1Dbm}
\end{figure}

Let us now consider the case of a relatively weak block mixing described by
$L_{SO}$.
Then the conductivity correction takes the form
\begin{eqnarray}
 \delta \sigma
  &=& \frac{e^2}{h}\left[ \left(
      \frac{W^4}{3L_H^4} f(W/l) + \frac{W^2}{L_\phi^2}
    \right)^{-1/2} \right.
    \nonumber
    \\
    &-& \left. \left(
      \frac{W^4}{3L_H^4} f(W/l) + \frac{W^2}{L_\phi^2} + \frac{W^2}{L_{SO}^2}
    \right)^{-1/2} \right].
    \label{1Dbm}
\end{eqnarray}
For a strong block mixing this equation reproduces the 1Sp result
(\ref{ballistic}),
while in the opposite limit of weak mixing the conductivity correction gets strongly
suppressed as illustrated in Fig. \ref{Fig1Dbm}.

%%%%%%%%%%%%%%%%%%%%%%%%%%%%%%%%%%%%%%%%%%%%%%%%%%%%%%%
%%%%%%%%%%%%%%%%%%%%%%%%%%%%%%%%%%%%%%%%%%%%%%%%%%%%%%%
%%%%%%%%%%%%%%%%%%%%%%%%%%%%%%%%%%%%%%%%%%%%%%%%%%%%%%%
%%%%%%%%%%%%%%%%%%%%%%%%%%%%%%%%%%%%%%%%%%%%%%%%%%%%%%%

\section{Summary and discussion}
\label{s6}

To summarize, we have reviewed manifestations of quantum interference in
conductivity of 2D HgTe structures. A
symmetry analysis yields a rich ``phase diagram''
 describing regimes with different types of one-loop quantum
interference correction (WL, WAL, no correction).
We have supplemented the symmetry analysis by a microscopic calculation of the
quantum interference contribution to the conductivity. This approach allows us
also to calculate the behavior of the conductivity in the crossover regimes
and beyond the diffusive approximation.
We have also discussed symmetry breaking mechanisms and the quantum
interference correction in a quasi-1D (strip) geometry.

%%%%%%%%%%%%%%%%%%%%%%%%%%%%%%%%%%%%%%%%%%%%%%%%%%%%%%%%%%%%%%
\begin{figure*}
 \includegraphics[width=0.95\textwidth]{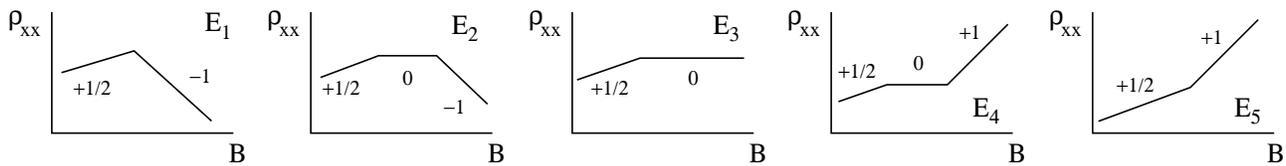}\hspace*{0.2cm}
 \caption{Schematic illustration of the magnetoresistivity $\rho_{xx}(B)$
 on a linear-logarithmic scale for energies
$E_1 \ldots E_5$ from the right panel of Fig. \ref{Fig1}.
The numbers denote the prefactors $\alpha=(\pi h/e^2) \partial \sigma/\partial
\ln \tau_H$ (here $\tau_H=L_H^2/D$) of the logarithmic magnetoresistivity.
}
\label{Fig2}
\end{figure*}
%%%%%%%%%%%%%%%%%%%%%%%%%%%%%%%%%%%%%%%%%%%%%%%%%%%%%%%%%%%%%%%%%%%%

The common way to explore the quantum interference (WL or WAL)
experimentally  is to measure the magnetoresistivity
$\rho_{xx}(B)$ in a transverse magnetic field $B$. Each of the symmetry-breaking
patterns discussed in Sec.~\ref{s2} then translates into a succession of regions
of magnetic field with corresponding signs and prefactors of the low-field
magnetoresistivity, see Fig.\ref{Fig2}.

Let us now discuss available experimental data in context of the above
theoretical findings. In Refs.~\onlinecite{kvon,minkov} the low-field
magnetoresistivity was
investigated in 2D HgTe samples both in the case of normal (thin QWs) and inverted
(thick QWs) band ordering. For both types of samples a weak
positive magnetoresistivity was observed in the lowest magnetic fields $B$.
This behavior can clearly be attributed to WAL.
The coefficient in front of $(e^2/\pi h) \ln \tau_H$
(where $\tau_H=L_H^2/D$) was found
to be consistent with  1/2, as expected for the 1Sp regime.
In higher magnetic fields, a crossover to negative magnetoresistance
was observed that could be presumably attributed to WL. Such a behavior
corresponds, in our terminology, to the $2\,{\rm O} \to 1\,{\rm Sp}$ symmetry pattern
(characteristic
for systems with relatively small carrier concentration, see regimes E$_1$ and E$_2$
in Fig.\ \ref{Fig2})\cite{footnote_ballistic}.  For inverted structures
with high carrier concentrations, Ref.\ \onlinecite{minkov} observed only positive
magnetoresistance consistent with regimes E$_4$ and E$_5$ of Fig.\ \ref{Fig2}.

In a recent work \cite{bruene} the magnetoresistance of a HgTe structure
patterned in an array of quasi-1D strips was experimentally studied. The WAL
behavior was observed both for normal-gap and inverted-gap structures in the whole
range of magnetic fields and gate voltages. This is consistent
with the 1Sp behavior as expected from the above theory in the quasi-1D regime.
Indeed, boundaries of the strips break the block-wise TR symmetry, as has
been shown in Sec.~\ref{s2.3}. Furthermore, diffuse boundary
scattering due to short-range edge irregularities is favorable for the
violation of the $z\to -z$ symmetry at the boundary, which breaks down the spin symmetry,
yielding the 1Sp regime.
In addition, the block mixing arises also due to the BIA, see
Ref.~\onlinecite{Qi11}, and due
to short-range impurities located within the quantum well. The difference in the
magnitude of the effect for normal-gap and inverted-gap setups observed in
Ref. \onlinecite{bruene} can be possibly related to a stronger block mixing
in the inverted case (cf. Fig. \ref{Fig1Dbm}) and/or stronger
dephasing in the normal case.  A stronger block-mixing in the inverted case might possibly be
attributed to the higher probability of having irregularities (background short-range impurities or edge roughness)
breaking the $z\to -z$ symmetry of the sample within a thicker QW (inverted band ordering) as compared to
a thin QW (normal band ordering).

Finally, it is worth emphasizing that, while we have focussed on HgTe/HgCdTe
quantum wells, the above analysis is expected to be applicable to a broader
class of structures with Dirac-type spectrum, including, in particular,
InAs/GaSb structures \cite{Liu08,Du}.

\section{Acknowledgments}

We are grateful to C. Br\"une, A. Germanenko, E. Hankiewicz, E. Khalaf, G. Minkov, L. Molenkamp, M. Titov,
and G. Tkachov for useful discussions. The work was supported by DFG-RFBR within DFG SPP ``Semiconductor
spintronics'', by DFG-SPP 1666, by RFBR, by grant FP7-PEOPLE-2013-IRSES of the EU network InterNoM, by GIF, and by BMBF.


\begin{thebibliography}{12}


\bibitem{Hasan10RMP}
M. Z. Hasan and C. L. Kane,
Rev. Mod. Phys. \textbf{82}, 3045 (2010).

\bibitem{Qi11} X.-L. Qi and S.-C. Zhang, Rev. Mod. Phys. \textbf{83}, 1057 (2011).


\bibitem{BernevigHughesZhang} B.A. Bernevig, T.L. Hughes, and
  S.-C. Zhang, Science \textbf{314}, 1757 (2006).

\bibitem{Koenig07} M. K\"onig, S. Wiedmann, C. Br\"une, A. Roth,
  H. Buhmann, L.W. Molenkamp, X.-L. Qi, and S.-C. Zhang, Science
  \textbf{318}, 766 (2007).

\bibitem{kane} C.\ L.\ Kane and E.\ J.\ Mele,
Phys.\ Rev.\ Lett. \textbf{95}, 146802 (2005); \textit{ibid.}
\textbf{95}, 226801 (2005).

\bibitem{Fu07}  L.\ Fu and C.\ L.\ Kane, Phys.\ Rev.\ B \textbf{76}, 045302 (2007);
L. Fu, C.L. Kane, and E.J. Mele, Phys. Rev. Lett.
\textbf{98}, 106803 (2007).


\bibitem{hasan} D. Hsieh, D. Qian, L. Wray, Y. Xia, Y.S.Hor, R.J. Cava, and M.Z.Hasan,
Nature \textbf{452}, 7190 (2008).

\bibitem{Roth09}
M.\ K\"onig, H. Buhmann, L.W. Molenkamp, T. Hughes, C.-X. Liu, X.-L. Qi, and S.-C. Zhang, J. Phys. Soc. Jpn. \textbf{77}, 031007
(2008); A.\ Roth, C. Br\"une, H. Buhmann, L.W. Molenkamp, J. Maciejko, X.-L. Qi, and S.-C. Zhang, Science \textbf{325}, 294 (2009).


\bibitem{Liu08} C. Liu, T.L. Hughes, X.-L. Qi, K. Wang, and S.-C. Zhang,
Phys. Rev. Lett. {\bf 100}, 236601 (2008).

\bibitem{Du}
I.\ Knez, Rui-Rui\ Du, and G.\ Sullivan,  Phys. Rev. Lett.  \textbf{107}, 136603 (2011);
L.\ Du, I.\ Knez, G.\ Sullivan, and Rui-Rui Du,  arXiv:1306.1925.

\bibitem{ostrovsky10} P. M. Ostrovsky, I. V. Gornyi, and A. D. Mirlin,
Phys. Rev. Lett. \textbf{105}, 036803 (2010).


\bibitem{Tkachov11} G. Tkachov and E. M. Hankiewicz, Phys. Rev. B \textbf{84},
035444 (2011).

\bibitem{OGM12} P. M. Ostrovsky, I. V. Gornyi, and A. D. Mirlin, Phys. Rev. B
\textbf{86}, 125323 (2012).

\bibitem{Richter12} V. Krueckl and K. Richter,  Semicond. Sci. Technol.
\textbf{27}, 124006 (2012).


\bibitem{McCann} E. McCann, K. Kechedzhi, V.I. Fal'ko, H. Suzuura, T. Ando,
and B.L. Altshuler, Phys. Rev. Lett. \textbf{97}, 146805 (2006);
K. Kechedzhi, E. McCann, V.I. Fal'ko, H. Suzuura, T. Ando,
and B. L. Altshuler, Eur. Phys. J. Special Topics \textbf{148}, 39 (2007).

\bibitem{aleiner-efetov} I. L. Aleiner and K. B. Efetov, Phys. Rev. Lett. \textbf{97}, 236801 (2006).

\bibitem{Ostrovsky06}
P.M. Ostrovsky, I.V. Gornyi, and A.D. Mirlin,
Phys. Rev. B \textbf{74},  235443 (2006).

\bibitem{Nestoklon} M.O. Nestoklon, N.S. Averkiev, and S.A. Tarasenko,
Solid State Comm. \textbf{151}, 1550 (2011).


\bibitem{Glazman} I. Garate and L. Glazman, Phys. Rev. B \textbf{86}, 035422 (2012).

\bibitem{Koenig13} E.J. K\"onig, P.M. Ostrovsky, I.V. Protopopov, I.V. Gornyi, I.S. Burmistrov, and A.D. Mirlin,
 Phys. Rev. B \textbf{88}, 035106 (2013).

\bibitem{minkov} G.M. Minkov, A.V. Germanenko, O.E. Rut, A.A. Sherstobitov, S.A. Dvoretski, and N.N. Mikhailov,
Phys. Rev. B \textbf{85}, 235312 (2012);
Phys. Rev. B \textbf{88}, 045323 (2013).

\bibitem{kvon} E.B. Olshanetsky, Z.D. Kvon, G.M. Gusev, N.N. Mikhailov, S.A. Dvoretsky, and J.C. Portal, JETP Lett. \textbf{91}, 347 (2010);
D. A. Kozlov, Z. D. Kvon, N. N. Mikhailov, and S. A. Dvoretsky, JETP Lett. \textbf{96}, 730 (2013).


\bibitem{bruene} M. M\"uhlbauer, A. Budewitz, B. B\"uttner, G. Tkachov,
E.M. Hankiewicz, C. Br\"une, H. Buhmann, and
L.W. Molenkamp,  Phys. Rev. Lett. \textbf{112}, 146803  (2014).



\bibitem{Rothe10} D.G. Rothe, R.W. Reinthaler, C.-X. Liu, L.W. Molenkamp, S.-C. Zhang, and E.M. Hankiewicz, New J. Phys. \textbf{12}, 065012 (2010).



\bibitem{Winkler} R.\ Winkler, \textit{Spin-Orbit Coupling Effects in
    Two-Dimensional Electron and Hole Systems},
Springer Tracts in Modern Physics, Volume 191, (Springer, Berlin, 2003).

\bibitem{footnote_mass} A similar role is played by disorder that acts differently on E and H subbands.
In the rotated basis used in Sec. \ref{s3}, the corresponding term in the Haniltonian is proportional to $\sigma_z s_z$,
similarly to $H_M$, and can be regarded as a random mass term. In the regime of an almost linear spectrum, $E_F\gg |m|$,
such disorder would lead to an additional contribution to $1/\tau_m$.
On the other hand, when the chemical potential is located near the band bottom, $E_F\ll |m|$, such type
of disorder does not yield additional symmetry breaking and only contributes to the elastic
scattering time $\tau$.




\bibitem{AA} B.\ L.\ Altshuler and A.\ G.\ Aronov, in {\em
    Electron-electron interactions
in disordered conductors}, edited by A.\ L.\ Efros and M.\ Pollak
    (Elsevier, 1985), p. 1.

\bibitem{schmid}  S. Chakravarty and  A. Schmid,  Physics Reports \textbf{140}, 4 (1986).

\bibitem{AAG}I.~L. Aleiner, B.~L. Altshuler, and M.~E. Gershenzon,
Waves Random Media \textbf{9}, 201 (1999).



\bibitem{comment}
In fact, this question is more subtle and the term $N=1$ requires special attention. It results in an ultraviolet logarithmic divergency (coming from  large momenta $q > k_F \gg 1/l$) of the return probability $W(\phi,\phi_0,0).$ This divergency leads to  ultraviolet renormalization \cite{aleiner-efetov}  of the key problem parameters (such  as $W_0$) and will not be discussed here. By default, we  assume that all parameters of the problem are already renormalized and  that $q\ll k_F.$


\bibitem{nonback}
A. P. Dmitriev, V. Yu. Kachorovskii, and I. V. Gornyi,
Phys. Rev. B \textbf{56}, 9910 (1997).

\bibitem{GKO14}
I. V. Gornyi, V. Yu. Kachorovskii, and P.M. Ostrovsky, Phys. Rev. B. \textbf{90}, 085401 (2014).



\bibitem{DK}
       V.K. Dugaev and D.E. Khmelnitskii,
Sov. Phys.  JETP \textbf{59}, 1038 (1984).

\bibitem{AltshulerAronov81}
B.\ L.\ Altshuler and A.\ G.\ Aronov, JETP Lett.\ \textbf{33}, 499 (1981).

\bibitem{Beenakker}
C.\ W.\ J.\ Beenakker and H.\ van Houten, Phys.\ Rev.\ B \textbf{38}, 3232
(1988).

\bibitem{AAK}
B.L.~Altshuler, A.G.~Aronov, and D.E.~Khmelnitskii,
J. Phys.\ C {\bf 15}, 7367 (1982).

\bibitem{footnote_ballistic}
A weak negative magnetoresistance in relatively strong fields may also arise in the E$_3$ regime of Fig.\ \ref{Fig2} due to ballistic effects, cf.\ Fig.\
\ref{FigII} (right panel).





\end{thebibliography}
\end{document}